\documentclass[10pt,journal,compsoc]{IEEEtran}

\usepackage{booktabs}
\usepackage{graphicx}
\usepackage{color}
\usepackage{array}
\usepackage{hyperref}
\usepackage{ragged2e}
\usepackage{enumitem}

\usepackage[utf8]{inputenc}
\usepackage{csquotes}
\usepackage[super]{nth}
 \usepackage[switch]{lineno} 

\usepackage[colorinlistoftodos,prependcaption]{todonotes}
\hypersetup{
    colorlinks,
    citecolor=blue,
    filecolor=blue,
    linkcolor=black,
    urlcolor=black
}

\newcommand{\myquote}[1]
{
  \vspace{1mm}
  \begin{quotation}
  \textit{``#1''}
  \end{quotation}
  \vspace{2mm}
}

\newcounter{EKXCommentsCounter}

\newcolumntype{L}[1]{>{\raggedright\let\newline\\\arraybackslash\hspace{0pt}}m{#1}}
\newcolumntype{C}[1]{>{\centering\let\newline\\\arraybackslash\hspace{0pt}}m{#1}}

\newlist{tablelist}{enumerate}{1}
\setlist[tablelist]{leftmargin=0.15in, label*=\arabic*),itemsep=0.02in}

\newcommand{\sparkline}[2]{(\raisebox{-.1\height}{\includegraphics[height=1em, width=5em]{#1}}, ``\textit{#2}'')}

\ifCLASSOPTIONcompsoc
  \usepackage[nocompress]{cite}
\else
  \usepackage{cite}
\fi

\ifCLASSINFOpdf
\else
\fi
\begin{document}
%

\title{A progression model of software engineering goals, challenges, and practices in start-ups}

\author{Eriks~Klotins,
        Michael~Unterkalmsteiner,
        Panagiota~Chatzipetrou,
        Tony~Gorschek,~\IEEEmembership{Member,~IEEE},
        Rafael~Prikladnicki,
        Nirnaya~Tripathi, and
        Leandro~Bento~Pompermaier

\IEEEcompsocitemizethanks{

\IEEEcompsocthanksitem E. Klotins, M. Unterkalmsteiner, and T. Gorschek are with the Software Engineering Research 
Lab Sweden, Blekinge Institute of Technology, Karlskrona, Sweden.

\IEEEcompsocthanksitem  P. Chatzipetrou is with Department of Informatics, CERIS, Örebro University School of Business, SE-701 82 Örebro, Sweden
and Software Engineering Research Lab, Blekinge Institute of Technology, Karlskrona, Sweden

\IEEEcompsocthanksitem  R. Prikladnicki and L. Pompermaier is with School of Technology at the Pontifical Catholic University of Rio Grande do Sul, Brazil.

\IEEEcompsocthanksitem  N. Tripathi is with M3S research unit, University of Oulu, Finland \protect\\

Corresponding author: Eriks Klotins, eriks.klotins@bth.se
}
\thanks{Manuscript received XX; revised XX}}

%
%

\markboth{Journal of \LaTeX\ Class Files,~Vol.~13, No.~9, September~2014}%
{Shell \MakeLowercase{\textit{et al.}}: Bare Advanced Demo of IEEEtran.cls for Journals}

\IEEEtitleabstractindextext{%
\justify
\begin{abstract}

\textit{Context:} Software start-ups are emerging as suppliers of innovation and software-intensive products. However, traditional software engineering practices are not evaluated in the context, nor adopted to goals and challenges of start-ups. As a result, there is insufficient support for software engineering in the start-up context.
\\
\textit{Objective:} We aim to collect data related to engineering goals, challenges, and practices in start-up companies to ascertain trends and patterns characterizing engineering work in start-ups. Such data allows researchers to understand better how goals and challenges are related to practices. This understanding can then inform future studies aimed at designing solutions addressing those goals and challenges. Besides, these trends and patterns can be useful for practitioners to make more informed decisions in their engineering practice.
\\
\textit{Method:} We use a case survey method to gather first-hand, in-depth experiences from a large sample of software start-ups. We use open coding and cross-case analysis to describe and identify patterns, and corroborate the findings with statistical analysis.
\\
\textit{Results:} We analyze 84 start-up cases and identify 16 goals, 9 challenges, and 16 engineering practices that are common among start-ups. We have mapped these goals, challenges, and practices to start-up life-cycle stages (inception, stabilization, growth, and maturity). Thus, creating the progression model guiding software engineering efforts in start-ups.
\\
\textit{Conclusions:} We conclude that start-ups to a large extent face the 
same challenges and use the same practices as established companies. However, 
the primary software engineering challenge in start-ups is to evolve multiple 
process areas at once, with a little margin for serious errors.


\end{abstract}

}

\maketitle

\IEEEdisplaynontitleabstractindextext

%
\IEEEpeerreviewmaketitle

\ifCLASSOPTIONcompsoc
	\section{Introduction}\label{sec:introduction} 
\else

\section{Introduction}
\label{sec:introduction}
\fi

Software start-ups are small companies created to build and market a software-intensive product~\cite{Sutton2000}.  
Start-ups are characterized by rapid evolution, small teams, uncertainty about customer needs and market conditions, and a high failure rate~\cite{Paternoster2014,berg2018software}. However, leveraging cutting-edge technologies, risk, and speed, start-ups can launch software products fast~\cite{Giardino2014,Baskerville2003}. 

Aims, objectives, and challenges of product engineering change quickly as a start-up evolves~\cite{Crowne2002}. State-of-the-art engineering methods offer little support for understanding the evolving context and selecting the right practices~\cite{Unterkalmsteiner,Paternoster2014}. A miscalculation in choosing engineering practices could lead to over or under-engineering of the product, and contribute to wasted resources and missed market opportunities~\cite{Giardino2014}.


According to industry reports, a record of 19.2 billion EUR of venture capital was invested in European and 80 billion USD in US start-ups in 2017 alone~\cite{Wijngaarde2018}. 
Building the first version of a product is a substantial engineering challenge and precedes any market or business related difficulties~\cite{Giardino2015,Crowne2002}. 
Thus, shortcomings in applied engineering practices could waste the investment, and hinder any subsequent attempts to market the product and to build a sustainable business around it. Even if a fraction of start-up failures could be attributed to engineering failures, that would still present an opportunity for better, start-up specific, engineering practices, and a relevant area of research.

An increasing number of studies attempt to explore engineering practices in a start-up context, for example, requirements engineering~\cite{Melegati2016, Gralha2018,tripathi2018anatomy}, technical debt~\cite{Klotins2018}, and user experience~\cite{Hokkanen2015}. Several studies, such as Giardino et al.~\cite{Unterkalmsteiner2014} and Crowne~\cite{Crowne2002}, attempt to explore and present conceptual models of product engineering in start-ups. However, none of these studies provide a full and detailed answer to what engineering practices start-ups use concerning different engineering process areas and start-up evolution stages. 
The need to better understand product engineering in start-ups, and to provide relevant support for practitioners, has been highlighted by Unterkalmsteiner et al.~\cite{Abrahamsson2016} and Klotins et al.~\cite{Klotins}.

With this study, we aim to understand how start-ups use different engineering process areas, and how utilized practices evolve over start-up life-cycle. 
Through our analysis, we present a progression model of what engineering 
aspects, that is, goals, practices, and challenges are relevant in start-ups in 
their evolution stage. The model is aimed at supporting practitioners in their 
decision making process and at pinpointing specific engineering challenges for 
further investigation.

We use an adapted case survey method~\cite{Petersen2017} to collect and analyze primary data on engineering practices in 84 start-up cases. The cases vary by geographical location, development stage, outcome, and time of operation among other factors, thus presenting a relatively large and diverse sample ~\cite{berg2018software}. 
To explore start-up goals, challenges, and used practices, we propose the 
start-up life-cycle model and analyze cases within the same development stage, 
and with the same outcome. We apply qualitative methods to identify patterns in 
the data and arrive at explanatory results. The explanatory results are 
verified and complemented with statistical analysis providing a firm basis for 
our conclusions.

Our study provides several novel contributions. Firstly, we present a start-up 
life-cycle model, aimed at illustrating the dynamically evolving nature of 
start-ups. Secondly, we use the life-cycle model to describe what practices, 
goals, and challenges are relevant to start-ups at different life-cycle stages. 
Thirdly, we present the start-up progression model aimed at guiding 
practitioners and at illustrating relevant areas for further exploration.

The remainder of the paper is structured as follows: in Section 2 we define 
software start-ups and summarize existing work in the area, in Section 3 we 
describe our research methodology, in Section 4 we report and analyze our 
results, in Section 5 we interpret and discuss our findings, Section 6 
concludes the paper.

\section{Background and related work}
\begin{figure*}[!th]
	\centering
	\includegraphics[width=\textwidth/10*9]{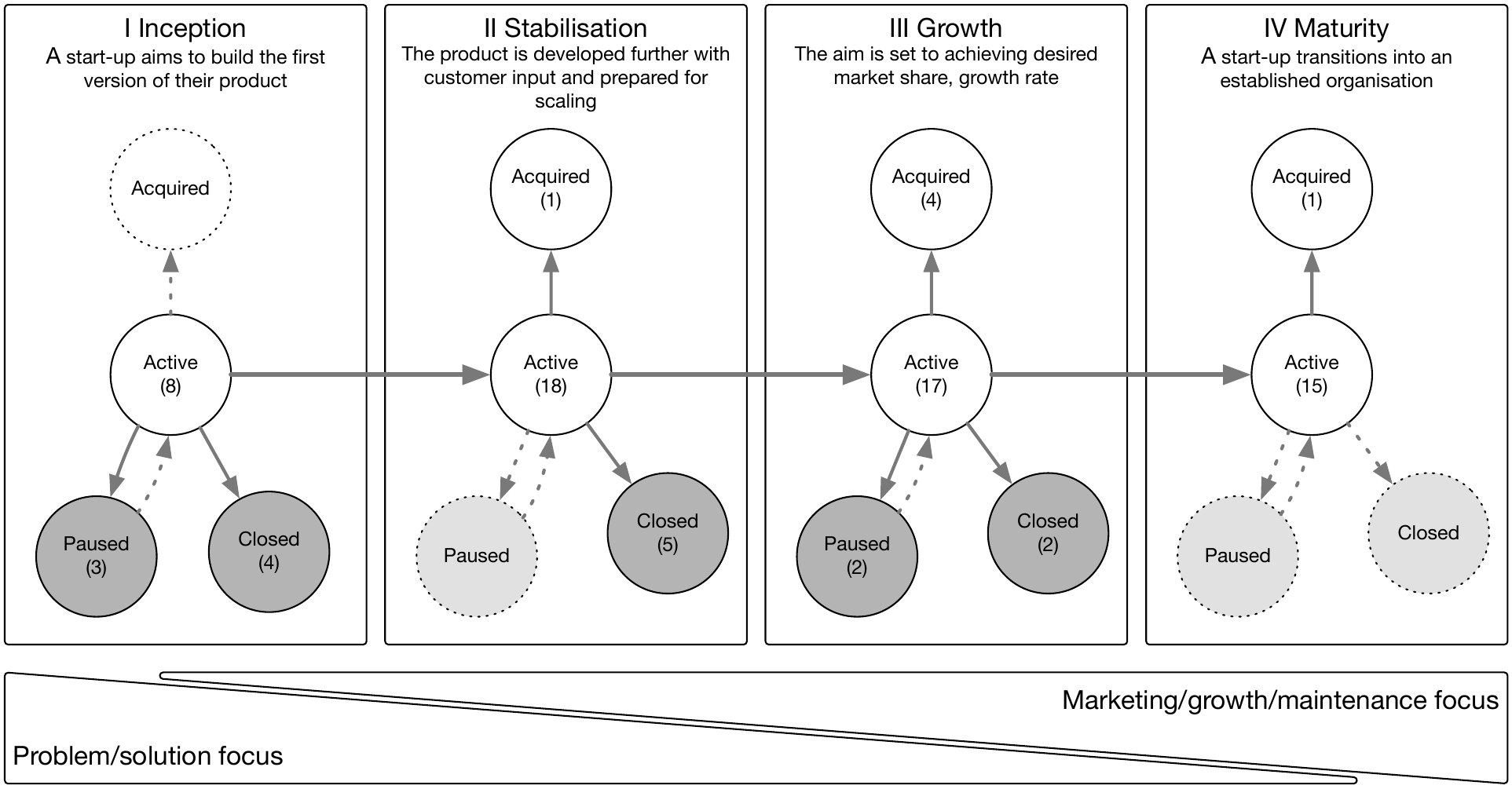}
	\caption{The start-up life-cycle model based on Crowne~\cite{Crowne2002} and Churchill et al.~\cite{NeilC.Churchill1983}. The white bubbles indicate ``good'' and desired states. The shaded bubbles show the undesirable states. Arrows denote possible transitions between the states. States and transitions that are possible, were however not observed in this study, are denoted by dashed lines. The numbers in brackets indicate how many cases representing each state were observed.}
  
	\label{fig_states}
\end{figure*}

\subsection{Software start-ups}

As early as 1994 Carmel~\cite{Carmel1994c} reported on small companies building and marketing innovative software products. 
These small companies, or start-ups, prioritize time-to-market over product quality, teams are small and self-motivated, and engineering practices informal. Later studies, for example, Giardino et al.~\cite{Giardino2014}, and Sutton et al.~\cite{Sutton2000} report the very similar characterization of start-ups.

Start-up companies are known for their high failure rate. About 75 - 99\% of start-up products fail to achieve any meaningful results in market~\cite{stateoftech2017,Blank2011}. The high failure rate could be explained by market challenges, team issues, difficulties in securing funding and so on. However, the capability to build software efficiently with limited understanding about stakeholder needs and with limited resources is the foremost challenge in software start-ups and precedes any market or business related challenges~\cite{Klotinsc}.

\subsection{What do we know about software start-ups?}

A broader interest to study software start-ups from software engineering perspective was launched by publication of a systematic review on existing literature in the area~\cite{Paternoster2014}. Selected results from this review were also published in IEEE Software~\cite{Giardino2014}. In the review, authors point out the potential of software start-ups as vehicles for innovation, and lack of relevant research in the area. The review lists a number of contextual challenges to software product engineering in start-ups compared to established companies. 

Two subsequent literature reviews were published by Klotins et al.~\cite{Klotins2015} and Berg et al.~\cite{berg2018software} aiming to map state-of-the-art in start-ups with SWEBOK~\cite{Society} knowledge areas. They conclude that there are many gaps and opportunities for developing start-up specific engineering practices. However, they attempt to analyze state-of-the-art in start-ups with SWEBOK that isn't start-up specific and lacks several relevant knowledge areas, for example, market-driven requirements engineering and value driven software engineering. 

Giardino et al.~\cite{Unterkalmsteiner} proposes the Greenfield start-up model identifying the main engineering categories and their relationships. The model identifies severe lack of resources, teamwork, rapid development, little focus to quality, evolutionary approach, and technical debt as the key categories. The relationships reveal that development speed is achieved by capable teams, little focus on quality assurance and internal product quality. However, such approach leads to accumulation of technical debt and hindered performance in the long term.


\subsection{Software engineering practices in start-ups}

Earlier studies point out that start-ups often do not follow best engineering practices and opt for an ad-hoc approach to engineering. Such an approach to engineering is partly due to the immaturity of start-ups, rapidly changing environment, and lack of engineering expertise. However, there is also limited support from academia pinpointing engineering practices that could be suited for such context~\cite{Klotins,Yau2013,Abrahamsson2016}. 

Part of the difficulty to practice and study software engineering in start-ups is the lack of knowledge transfer between start-ups. All knowledge about a domain, ways-of-working, and practices is carried by individuals, thus is lost when a start-up closes down, and needs to be reinvented with every new start-up. Thus, a large part of start-up research is to establish a body of knowledge with the best engineering practices for start-ups~\cite{leonard1995wellsprings,Klotins}.

Several studies attempt to explore product engineering practices in start-ups. For example, Gralha et al.~\cite{Gralha2018} and Melegati et al.~\cite{Melegati2016} explore requirements engineering practices in start-ups. These studies suggest that requirements engineering is one of the key engineering process areas in start-ups, and help to explore the market opportunity and to devise a feasible solution.

Hokkanen et al.~\cite{hokkanen2016minimum} present a framework guiding user experience design work in a start-up context. They argue that for a product to be successful in a market it has to fulfill minimal functional and user experience requirements. The framework identifies and prioritizes various user experience elements guiding user experience focus.

In our earlier work (Klotins et al.~\cite{Klotins2018}) we explore technical debt in start-ups. We found that excessive technical debt could be a cause for missed market opportunities and contribute to start-up failures. Furthermore, they identify several strategies that could help to expose unwanted technical debt early.

The earlier work suggests that engineering practices are rapidly evolving to accommodate the changing engineering context~\cite{hokkanen2016minimum,Gralha2018,Unterkalmsteiner,Klotinsc}. Even though, individual areas, for example, requirements engineering and user experience design, has been studied, there is a lack of a coherent view of how different engineering areas fit and evolve together. Such complete understanding is needed to analyze why specific practices are applied, and to spot potential misalignment between engineering process areas.

\subsection{Start-up life-cycle models}\label{sec_model}

To illustrate the changing objectives of a start-up, there have been attempts to define start-up life-cycle models. Blank~\cite{Blank2013} identifies two stages, search for a viable market opportunity, and building a viable business around the opportunity. Each stage consists of several objectives outlining the need to explore and validate a potential customer need first, then propose and validate a potential solution, and finally scale up the operations. This model, however, is generic and offers little guidance for software engineers.

Inspired by Churchill et al.\cite{NeilC.Churchill1983} and 
Crowne~\cite{Crowne2002} we combine start-up product evolution stages with 
relevant organization states (Fig.~\ref{fig_states}). In this paper, we use the 
four stages of a start-up as a basis for explaining start-up evolution and 
define them as:

\begin{enumerate}
  \item Inception - a stage between ideation of a product until the start-up releases the first product release to the first customer. The primary goal of this stage is to scope and build the minimum viable product by balancing needs of a customer, available resources, and time~\cite{Junk2000}.

  \item Stabilization - a stage between first product release until readiness for scaling. In this stage, a start-up aims to ensure that the product can be decommissioned without adding extra effort to the development team. That is, the product should be easy to maintain, scale, and surrounding infrastructure, for example, operations and customer support, are in place.
  
  \item Growth - at this stage the focus is set on attaining the desired market share and growth rate. Although the efforts shift towards marketing and sales, the engineering team should cope with a flow of new customer requirements, and product variations for different markets.

  \item Maturity - in this stage a start-up transitions into an established organization aiming to 
  preserve established market share and to optimize its operations. The engineering team should install routines for operating and maintaining the product.

\end{enumerate}

These four stages represent the shift in start-up objectives. Early stages focus on finding a relevant problem, devising a feasible solution. Later, the focus shifts to marketing and improving the efficiency of start-up's operations~\cite{Blank2013b, Crowne2002}. 

In case a start-up decides to radically change some fundamental aspect of its product, that is, to pivot~\cite{bajwa2016software}, the product likely moves to an earlier phase in the life-cycle model. For example, discarding existing features and developing new ones implies abandoning any marketing or stabilization efforts related to the abandoned features. Developing entirely new features throws the start-up back to inception stage for scoping, validation and piloting.

At any of the four product stages a start-up strives to: a) remain active and continue operations (that is, not to fail), b) advance to the next stage or c) be acquired by another company for profit to shareholders. Alternatively, at any of these stages, a start-up may be closed down or paused. We define organization states as the following:

\begin{enumerate}
  \item Active - the team actively works on the product.

  \item Paused - the team has stopped working, however there is an intention to resume at some point in future. Reasons for pausing a start-up could be, for example, a temporary shift in founders priorities, temporary lack of funding for product development or marketing among other scenarios. 

  \item Acquired - another company acquires the start-up for a profit to shareholders. The team is disbanded or merges with the other company.

  \item Closed - the team is disbanded or works on something else.
\end{enumerate}

The combined model of start-up and product life-cycle states is shown in 
Fig.~\ref{fig_states}. We use this model as an input to our study and to capture the state of each studied case.


\section{Research methodology}
\begin{table*}[!t]
	\renewcommand{\arraystretch}{1.5}
	\caption{Steps of the research method}
	\label{table_methods}
	\centering
	
	\begin{tabular}{L{0.2in}L{1in}L{1.8in}L{3in}}
		
		Step & Case survey method (Larsson~\cite{Larsson1993}) & Theory 
		building (Eisenhardt~\cite{Eisenhardt1989}) & The applied method \\
		\hline
		1 & - & Define research questions and any a priori constructs & We 
		start the study with a broad aim to collect primary data on how 
		start-ups practice software engineering. We define our research 
		questions in Section~\ref{sec_research_questions}, and present the 
		start-up life-cycle model in Fig.~\ref{fig_states}.
		\\
		
		2 &  Select cases of interest & Specify a population & We aim to study 
		start-up companies building software-intensive products and reach as 
		diverse sample of start-ups as possible. \\

		3 & Design the data extraction form for elicitation & Craft instruments 
		and protocols for data collection & 
		We design a questionnaire for data collection. It is aimed at start-up 
		practitioners with practical experience with software engineering in 
		start-up setting, see Section~\ref{sec_questionaire_design}. \\
		
		4 & - & Parallelize data collection and analysis & We parallelize the 
		final steps of questionnaire design with the data collection to 
		identify any issues with the questionnaire before scaling up the data 
		collection. \\
		
		5 & Conduct the coding & Conduct within case analysis, search for 
		cross-case patterns using divergent techniques & 
		First, we use open coding to extract relevant information from textual 
		data. Secondly, we use the start-up life-cycle model, see 
		Fig.~\ref{fig_states}, to categorize the cases and conduct cross-case 
		analysis, see Section~\ref{sec_crosscase_analysis}. \\
		
		6 & Use statistical approaches to analyze the coding & 
		Iterative shaping of hypotheses, search evidence for ''why" behind 
		relationships. 
		& We condense findings from the previous step into more broader 
		patterns, and perform a parallel quantitative analysis complementing 
		qualitative results, see Section~\ref{sec_patterns}.\\
		
		7 & - & Comparison with conflicting and similar literature & We compare 
		patterns from the previous step with literature. We seek for additional 
		support and explanation for our findings, see 
		Section~\ref{sec_patterns}.  \\
		
		8 & - & Aim for theoretical saturation when possible & -

		\\

		\hline

	\end{tabular}
\end{table*}

We use a case survey method to collect and analyze primary data from start-up companies. The method is aimed to balance studying a few cases in depth with traditional (multi) case studies and quantitatively studying many cases~\cite{petersen2017choosing}. Case studies offer greater level of detail and internal validity by closely examining multiple data sources. Surveys attempt to collect data from a large number of cases, thus achieving a higher degree of generalizability~\cite{Runeson2012}. 

The main advantages of the case survey method are its ability to collect richer information than conventional survey, and extendibility to study more cases. That said, the data collected by a case survey are limited by the scope of the questionnaire instrument\cite{Larsson1993,petersen2017choosing}.  Validity threats of our study are discussed in Section~\ref{sec_threats}.

The original case survey method description suggests using coding and 
statistical methods to analyze the data~\cite{Larsson1993}. However, we extend 
the method by adding more steps from the theory building process proposed by 
Eisenhardt~\cite{Eisenhardt1989}. While both methods are compatible,  
Eisenhardt provides more details on inducing a theory from data, that is, in 
this study, the start-up progression model. We present a mapping between the 
two methods and the combined method in Table~\ref{table_methods}.

\subsection{Research questions}\label{sec_research_questions}

To guide our study, we define the following research questions (RQ):
\\
\\\textbf{RQ1:} What patterns pertaining software engineering can be ascertained in start-up companies?
\\\textit{Rationale:} Very little is known of what software engineering 
practices start-ups use and what is the motivation for using or avoiding 
specific practices. Earlier studies report the use of light-weight, ad-hoc 
practices with emphasis on requirements 
engineering~\cite{Klotins2015,Paternoster2014,Melegati2016}. However, most 
earlier reports use secondary data, explore only a few cases, and are based on limited understanding of engineering context in start-ups. 

With this research question, we identify commonalities in engineering goals, 
challenges, and used software engineering practices in start-ups with respect 
to their life-cycle stage, see Fig.~\ref{fig_states}. An understanding of what 
goals and challenges shape the use of engineering practices are essential to: 
\begin{enumerate}[label=\alph*)]
  \item Judge suitability of commonly used practices.
  \item Devise new engineering practices to navigate specific challenges and to 
  achieve particular goals.
  \item Provide a blueprint of software-intensive product engineering in a 
  start-up context~\cite{Klotins}.

\end{enumerate}

We formulate three sub-questions to explore goals, challenges, and practices specifically.
\\
\\\textbf{RQ1.1:} What goals, relevant to software engineering, can be 
ascertained in start-up companies?
\\\textit{Rationale:} We aim to explore what goals are driving software 
engineering in start-ups concerning their life-cycle stage, see 
Fig.~\ref{fig_states}. A fine-grained understanding of the goals, i.e. drivers 
for engineering activities, helps to understand the  context of why certain 
engineering practices are used, or avoided~\cite{Kirk2014a,Klotins}. 
\\
\\\textbf{RQ1.2:} What challenges relevant to software engineering can be ascertained in start-up companies?
\\\textit{Rationale:} Engineering challenges is another context factor, alongside goals, shaping engineering practices in start-ups. We aim to explore what specific challenges, associated with start-up life-cycle stages, can be ascertained in start-ups.
\\
\\\textbf{RQ1.3:} What software engineering practices do start-ups use?
\\\textit{Rationale:} We aim to explore what engineering practices start-ups apply as a response to life-cycle stage-specific goals and challenges.

\subsection{Data collection and analysis}

\subsubsection{Questionnaire design}\label{sec_questionaire_design}


We base the data extraction on a questionnaire eliciting practitioner experiences about their specific start-up case.  The scope of the survey is inspired by our earlier work the Start-up Context Map~\cite{Klotins} and covers team, requirements engineering, value, quality assurance, architecture and design, and project management aspects of start-ups.

During the questionnaire design, we conducted multiple internal and 
external reviews to ensure that all questions are relevant, easy to understand and that we receive meaningful answers. 

The internal reviews were conducted among the authors to determine the scope and flow of questions. For the external reviews we invited 10 members of Software Start-up Research Network\footnote{Software Start-up Research Network: \url{http://softwarestartups.org}} to provide their input. Firstly, we asked them to fill in the questionnaire and answer all the questions as if they were engineers in a start-up. Then, we organized a joint on-line workshop to discuss participants reflections and potential improvements to the questionnaire. Their responses were removed from the final dataset.

Finally, we piloted the questionnaire with four practitioners from different 
start-ups. During the pilots, respondents filled in the questionnaire while 
discussing questions, their answers and any potential issues with the first author of 
this paper.

As a result of these reviews, we improved the question formulations and removed 
some irrelevant questions. The finalized questionnaire contains, 10 sections, 85 high level 
questions and 285 sub-questions, 73 of the sub-questions are free-text, others are on an ordinal or nominal scale. Note that one question in the questionnaire may result in multiple sub-questions, for example, a high level question may result in two sub-questions one capturing a Likert-scale answer, another free text motivation for the answer. The sections cover many software engineering topics, see Table~\ref{table_topics}. Note that through the analysis process some topics were merged, and some lifted out. The last column of the table, process area, shows a mapping between sections of the questionnaire and the resulting process areas. 

From all the questions, 54 sub-questions focus on capturing respondents agreement with statements addressing their engineering practices and engineering context. To capture respondents level of agreement with a statement we use a Likert scale: not at 
all (1), a little (2), somewhat (3), very much (4). 
The values indicate the degree of agreement with a statement. Statements are formulated consistently in a way that lower values indicate less agreement, however higher values indicate more agreement. We specifically avoid neutral (neither agree or disagree) answer option to force respondents to state their opinion. However, we provide the ``I do not know'' option to allow respondents to explicitly skip the question.

All the questions and answer options are available as supplemental material\footnote{Full questionnaire:

\url{http://eriksklotins.lv/files/GCP_questionnaire.pdf}}.


\begin{table}[!ht]
\renewcommand{\arraystretch}{1.5}
\caption{Topics covered by the questionnaire}
\label{table_topics}
\centering

\begin{tabular}{L{0.11in}L{1.1in}C{0.5in}C{0.5in}C{0.4in}}

\# & Topics & Questions & Number of sub-questions & Process area\\
\hline
1 & Start-up demographics & 1 - 8 & 12 & -  \\
2 & Product demographics & 9 - 17 & 18 & - \\
3 & Respondent demographics & 18 - 23 & 10 & I \\
4 & Team demographics  & 24 - 30 & 12 & I \\
5 & Requirements engineering & 31 - 47 & 57 & II, III \\
6 & Software architecture & 48 - 55 & 19 & V \\
7 & User interface design & 56 - 59 &  9 & V \\
8 & Development practices & 60 - 69 & 80  & I - VI \\
9 & Quality assurance & 70 - 77 & 48 & IV \\
10 & Project management & 78 - 85 & 20 & VI \\

\hline
& Total & & 285 \\
\end{tabular}
\end{table}

\subsubsection{Distribution of the survey and data collection}\label{sec_sampling}

To distribute the survey we reached out to The Software Start-up Research Network\footnote{Software Start-up Research Network: \url{http://softwarestartups.org}} and asked other researchers to use their networks and connections. All authors of this paper actively promoted the survey through their on-line accounts and personal contacts. The first author promoted the study in several start-up themed events. The data collection took place between December 1, 2016 and May 15, 2017.

To support the data collection we created an on-line tool. The landing page of the tool contained a short description of the study aims, and a promotional video encouraging start-ups to share their experiences. The questionnaire was public and freely available to everyone. To screen the responses and gauge their suitability for our study, the questionnaire contains a multitude of demographical questions about the start-up and the respondent. For example, what is respondents relationship with the start-up, their role in the company, and how long time ago they were in contact with the start-up?

A total of 369 practitioners started to answer the questionnaire. However, many of the responses were incomplete. We removed responses with less than 50\% of questions answered. We further removed several duplicates and responses from non-software start-ups. The response rate of the questionnaire was 23\%, 84 out of 369.

\subsubsection{Coding scheme and cross-case analysis}\label{sec_crosscase_analysis}

We perform the cross-case analysis using both qualitative and quantitative methods. 

The responses are already structured by questions, and for multiple-choice questions, responses are already categorized. Thus, we need to code only responses from open-ended questions. Such questions help us to gain a finer understanding on how each topic is implemented in start-ups. Since there is no established body of knowledge of software engineering in start-ups, we use open, in-vivo, coding to gain insights how respondents themselves perceive and use software engineering. Thus, we applied open coding to identify described stakeholders, practices, artifacts used or produced, steps taken, and motivations for certain decisions~\cite{Corbin1990}. Each open-ended question addresses a different topic, thus we developed an individual coding scheme for each question.

The questionnaire is formulated to capture both used practices, and the practitioners' experiences with using the practices. Respondent reflections were captured in separate questions formulated along the lines of: ``In hindsight, what would you have done differently?''. In our results, we report and describe applied practices and respondents' experiences with the practices separately. However, the progression model in Fig.~\ref{fig_tpa-overview} contains only practices that were reported as used.


We use the start-up life-cycle model, see Fig.~\ref{fig_states}, to group cases by start-up life-cycle state. We analyze start-ups within each group and look for recurring engineering practices, challenges, goals, and contextual factors in their responses. We document these findings with  memos.
A memo describes a finding, start-up cases that were basis for the finding, category of the finding, that is, whether it pertains goals, practices, context factors or challenges, and to what state of the start-up life-cycle model the finding pertains to. We developed a total of 1856 codes and 366 memos. An example of open coding and memos is available as supplemental material on-line\footnote{Example of the open coding:\\\url{http://eriksklotins.lv/files/GCP_open_coding.pdf}}.


The initial coding is performed by the first author. The resulting memos are discussed and refined by the first and the second author jointly. In this process, memos with limited support from the data are removed, or merged with other similar memos. Interesting analysis points were marked for additional analysis. This is an iterative process leading to formulation of our findings.

As the final steps of our analysis, we use descriptive statistics and contingency tables to identify new and to confirm already identified patterns~\cite{haberman1973analysis}. We illustrate our results with frequency analysis and histograms, and test statistical significance of our findings.

We use the Chi-Square test of statistical association to test if the associations between the examined variables are not due to chance. To prevent Type I errors, we used exact tests, specifically, the Monte-Carlo test of statistical significance based on 10~000 sampled tables and assuming $(p < 0.05)$~\cite{hope1968simplified}. 

To examine the strength of associations we use Cramer's~V test. We interpret the test results as suggested by Cohen~\cite{cohan1988statistical}. That is, we consider thresholds of $0.1, 0.3,$ and $0.5$ for weak, moderate and strong associations.




To explore the specifics of an association, such as which cases are responsible 
for this association, we perform post-hoc testing using adjusted residuals. We 
consider an adjusted residual significant if the absolute value is above 1.96 
$(Adj.residual > 1.96)$, as suggested by Agresti~\cite{agresti1996introduction}. 

The adjusted residuals drive our analysis on relationships between start-up 
demographics and reported engineering practices. However, due to the 
exploratory nature of our study, we do not state any hypotheses upfront and 
drive our analysis with the research questions. We document statistically 
significant findings with field memos for further analysis.

\subsubsection{Development of patterns}\label{sec_patterns}

To develop our results further, we revise and group together similar memos from the previous step to formulate broader findings. This process is aimed at building up evidence for supporting a particular finding. As suggested by Eisenhardt~\cite{Eisenhardt1989}, we apply the following practices:

\begin{itemize}
  \item We analyze outlying, extreme, or otherwise surprising findings to shape our findings.
  \item We collect multiple variables on each topic and seek to triangulate our findings with multiple variables, and cases. 
  \item We search for cases that present contradictory evidence to our propositions and shape our propositions to cover the negative evidence.
  \item To decide if our findings constitute a salient pattern we use a combination of criteria. Firstly, we look if a finding appears in more than one case. Secondly, we look if multiple variables support the finding, in particular, whether respondents have pointed it out in their reflections. Thirdly, if the finding is supported by statistical analysis.
\end{itemize}

As a result of this step, we identify 55 patterns pertaining to software engineering in start-ups. Similar to the memos before, a pattern describes a specific practice, challenge, context factor, or goal, along with information what cases were the basis for formulating this pattern, and information in what context this pattern was observed. The patterns are further grouped into 6 engineering areas and categorized into 33 patterns illustrating goals, practices, and challenges.


Goals are patterns describing a desirable outcome toward which an effort is directed. Such desirable results are identified either explicitly by question formulation, e.g. ``What is the primary quality goal?'', or by the practitioners' own reflections on why a certain practice was used. In few occasions, a goal overlaps with a practice, e.g. to use product usage metrics to gauge start-up performance, see G16 and P14 in Fig.~\ref{fig_tpa-overview}. Such overlap indicates an association between use of the practice and attainment of the goal.

Practices are engineering activities helping start-ups to advance through the life-cycle stages. A practice could be means-end, such as establishing a team, could help to solve a problem, e.g. by documenting feature ideas, or help to gather knowledge on the current needs, e.g. by determining ``good-enough'' level of quality~\cite{sheppard2007engineering}.   

Challenges are difficulties in practicing software engineering and progressing through the life-cycle stages. We identify challenges from respondents reflections on how practices were applied and what they would do differently the next time. Some challenges overlap with practices, e.g. to establish a feedback loop with customers, see for example, C4 and P5 in Fig.~\ref{fig_tpa-overview}. Such overlap indicates an association between the practice and the challenge, i.e. that start-ups attempt to use a specific practice, however find it challenging.

As a result of this step, we identify patterns pertaining to software 
engineering in start-ups. Similar to memos before, a pattern describes a 
specific practice, challenge, context factor, or goal, along with information 
what cases were the basis for formulating this pattern, and information in what 
context this pattern was observed. The patterns are further grouped into 
engineering process areas, resulting in the start-up progression model.




\subsection{Threats to validity}\label{sec_threats}

Larsson~\cite{Larsson1993} identifies a number of validity concerns for case survey research.

\textit{Descriptive validity, factual accuracy} could be compromised if responses from start-ups lack important information leading to an incorrect interpretation of the cases. To address this threat we iterated the questionnaire instrument with both researchers and start-ups to assure that all important aspects of software engineering in start-ups are covered and our questions are understandable to practitioners.
Furthermore, we provided an explicit option to capture ``I do not know'' answers and, at the end of each section, asked ``What would you do differently next time?'' question to capture practitioner reflections on the most important lessons learned.



\textit{Respondent bias} stems from participants' inability or unwillingness to provide accurate responses. 

We collect respondents experiences and estimates of events that may have occurred in the past. Thus, the quality of the responses depend greatly on respondents memory and ability to reconstruct past events. Respondent responses suggest that majority of them, 67 out of 84, 80\%, are currently involved with their start-ups or have been involved in the last 12 months. Thus, the majority of responses concern relatively fresh experiences.

Some of the respondents, 36 of 84, 43\%, have specified that their main area of expertise in their start-up is other than software engineering, see Fig.~\ref{fig_respondent_bg}. There could be a concern whether they can provide reliable answers to questions about software engineering. Earlier work suggests that start-up teams are closely-knit, team members perform multiple roles, and the effort is focused on launching one product~\cite{Unterkalmsteiner,Paternoster2014}. Thus, even if a respondents primary area of expertise is not software engineering, they are closely involved in product work.

We further explore potential differences in answers due to time since the last contact with the start-up and respondents area of expertise with statistical tests. As part of the last step of cross-case analysis, see Section~\ref{sec_crosscase_analysis}, we test if respondents association with the start-up, age, time since the last contact, area of expertise, amount of experience in the area of expertise, and amount of experience working with start-ups, has any effect on their responses. Significant results from this analysis are presented in respective process areas. 

There is a possibility that respondents are unwilling to provide honest responses, or twist the responses to what they believe researchers hope to hear. In the particular case of startups, and especially when respondents are answering about a failed case, it may be hard for them to admit were they failed and if they were not following what is perceived as the best practices in software development.

Participation in the study was voluntary, and we advertised the questionnaire as means to help other start-ups by sharing what worked and what did not in their start-ups. Thus, we minimize biases stemming from participants being forced to participate and to provide dishonest answers.
The questionnaire contains a mix of multiple-choice, Likert scale, and free text providing multiple means of capturing the experiences, and mitigating the chance of extreme responses. We offer ``What would you do differently next time?'' questions to capture respondents own lessons learned.

\textit{Interpretive validity, objectivity of the researcher} is concerned with potential bias stemming from researchers misinterpreting the data. To address this threat, the first three authors of this paper frequently met and discussed any intermediate findings as part of the analysis process. As a result, findings with weak support from the data were identified and removed, see Steps 4-6 in Table~\ref{table_methods}.

\textit{Generalizability} of our results is determined by the number and diversity of the studied cases. We explore a diverse set of 84 start-up cases varying by geographical location, domain, product life-cycle stage, the extent of team expertise, and start-up outcome. That said, operational companies are overrepresented in our sample, see Fig.~\ref{fig_timeline}, and could bias our conclusions towards active, that is, to some extent successful start-ups. To compensate for this potential bias, we analyze operational and closed companies separately, compare the results, and apply statistical methods to determine significance of any conclusions.

Our sample mostly contains start-ups from Europe and South America, and start-ups from North America and Asia are underrepresented. For example, start-ups in the U.S. may have access to a broader and more homogeneous market than their European counterparts who need to adapt their products to the diversity of Europe's markets. Such differences could have an effect on software engineering goals, challenges, and practices.

\textit{Repeatability} of case surveys increases the objectivity of the findings. To strengthen repeatability of this study, we provide the data extraction form, full demographical information of the studied cases, and raw data\footnote{Upon request to the first author, 
Eriks Klotins, \url{eriks.klotins@bth.se}} as supplemental material.





\section{Results and analysis}


After removing incomplete and irrelevant responses, we analyze 84 responses from start-ups building software-intensive 
products. 

\begin{figure}[!t]
\centering
\includegraphics[width=\textwidth/2]{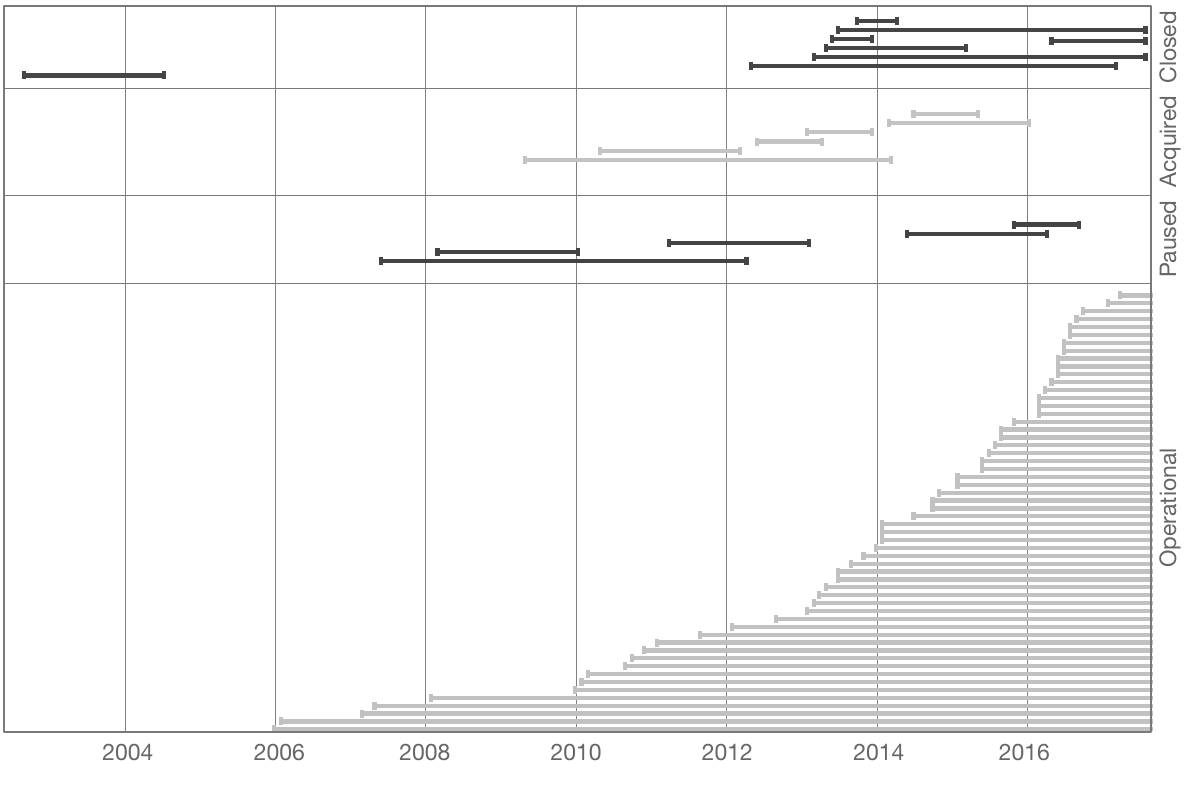}
\caption{Gantt chart illustrating operation time and outcome for studied companies. 9 respondents had not answered when they started working on the product, thus these cases are not shown in the figure}
\label{fig_timeline}
\end{figure}

Some of the first questions in the questionnaire collects demographical information about the start-up, such as current state, state of the product, when the team started working on the product. The responses show that our sample consists of start-ups established between 2004 and 2017, 
with the majority (60 of 84, 71\%) of start-ups actively working on their 
products at the time of the survey; 24 are closed, paused or acquired by other companies, see 
Fig.~\ref{fig_timeline}. 

Responses on the product state show that our sample contains start-ups in all life-cycle stages, inception - 15 (18\%), stabilization - 24 (29\%), growth - 26 (30\%), 
and maturity - 15 (18\%), however 4 companies have not specified their product state. Most 
start-ups have their main office in South America (39 of 84, 46\% ) and Europe 
(33 of 84, 39\%), the rest are located in Asia and North America. Such underrepresentation of North America and Asia could be explained by origin of the authors. The authors represent Europe and South America and were actively promoting the study in their networks.

\begin{figure}[!t]
\centering
\includegraphics[width=\textwidth/2]{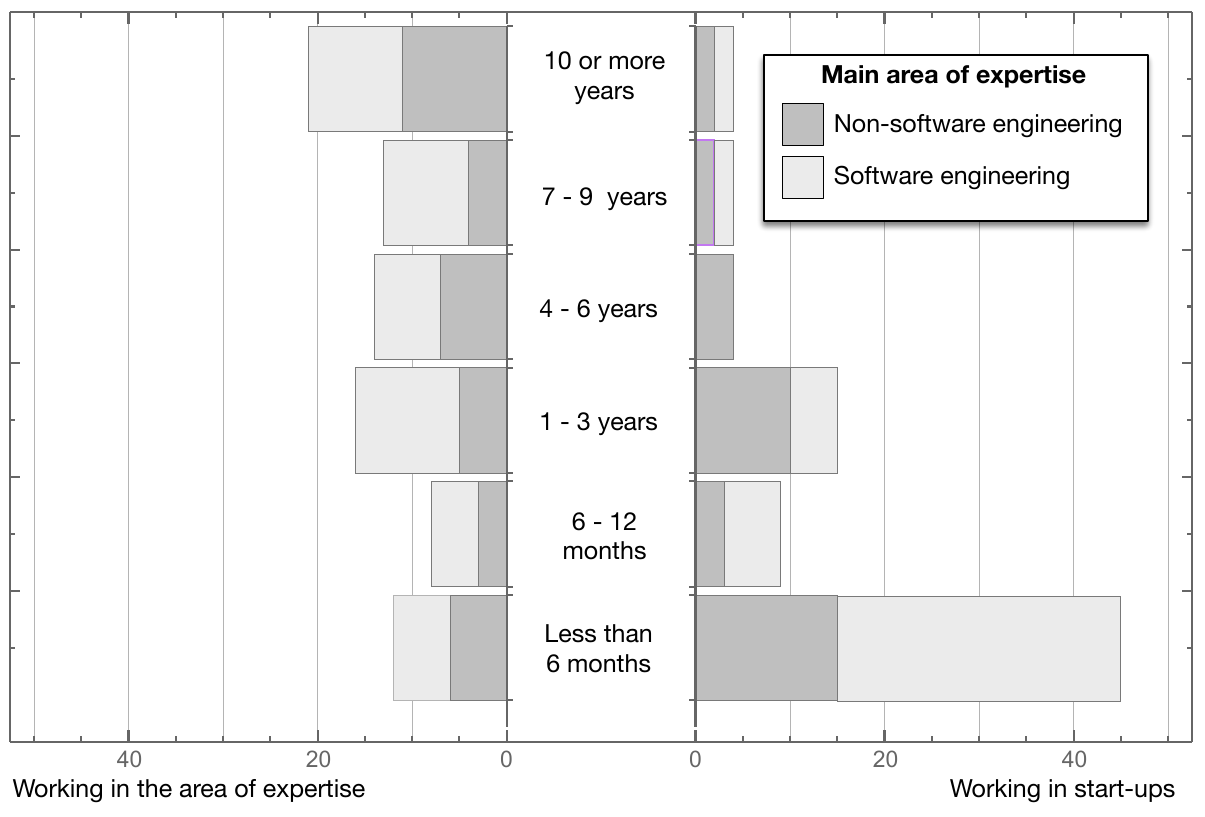}
\caption{Distribution of respondent background and prior experience}
\label{fig_respondent_bg}
\end{figure}

With the questionnaire we collect respondents demographical information, such as age, experience, area of expertise, relationship with the start-up, and how recent they have been involved in the start-up. Responses show that most of the respondents (62 of 84, 74\%) are founders, others are employed by start-ups (16 of 84, 19\%), or are otherwise associated. The respondents are about evenly distributed regarding whether their area of expertise is software engineering or not, an how much prior experience they have in their area of expertise (left side of Fig.~\ref{fig_respondent_bg}). However, most have little experience with start-ups before the current case (right side of Fig.~\ref{fig_respondent_bg}).

Other details characterizing our sample and engineering context in start-ups are
presented along with their respective process areas, discussed in the remainder of
this section. A list of studied cases, respondents, and their demographical 
information is available as supplemental material\footnote{All cases and their 
demographical information:

\url{http://eriksklotins.lv/files/GCP_demographics.pdf}}.

\begin{figure*}[!t]
\centering
\includegraphics[width=\textwidth]{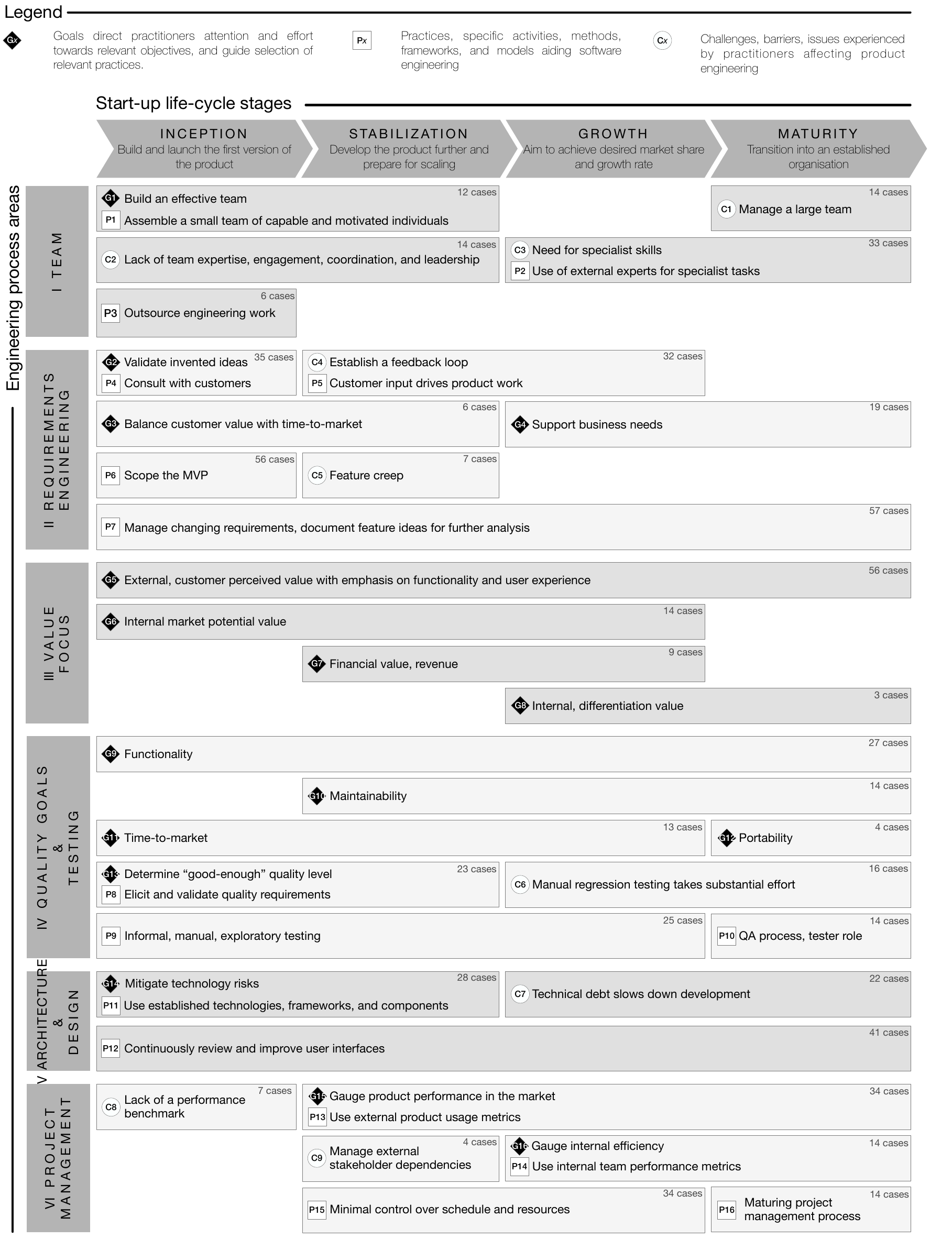}
\caption{The start-up progression model outlining software engineering goals, challenges, and practices in start-ups}
\label{fig_tpa-overview}
\end{figure*}

We structure our results into 6 process areas: team, requirements engineering, value, quality assurance, architecture and design, and project management, see Table~\ref{table_topics}. Within each process area we consider goals, challenges, and practices, i.e., engineering aspects, see the legend in Fig.~\ref{fig_tpa-overview}. 

We analyze the process areas in relation to the four start-up evolution stages, inception, stabilization, growth, 
and maturity. In Fig.~\ref{fig_tpa-overview} we provide an overview of the results and present the start-up progression model. 
To maintain traceability between our description and the figure we enumerate our findings in the following way: 
goals (\textit{Gx}), challenges (\textit{Cx}) and practices (\textit{Px}). On the top-right corner of each bar we denote how many cases were basis for formulating the finding.

The purpose of the progression model is to provide an overview of the critical 
stages of product development, main concerns, and relevant practices. 
For researchers, the model summarizes the key engineering concerns for further investigation and provides a structure for adding new results.
For practitioners, the model helps to identify the current product stage, the key objectives to be attained and lists the key engineering practices for attaining said objectives. We discuss each process area, goal, challenge and practice in the following sections.

To present respondent estimates on Likert scale questions, we illustrate the distribution of answers with in-line histograms, for example,
~\sparkline{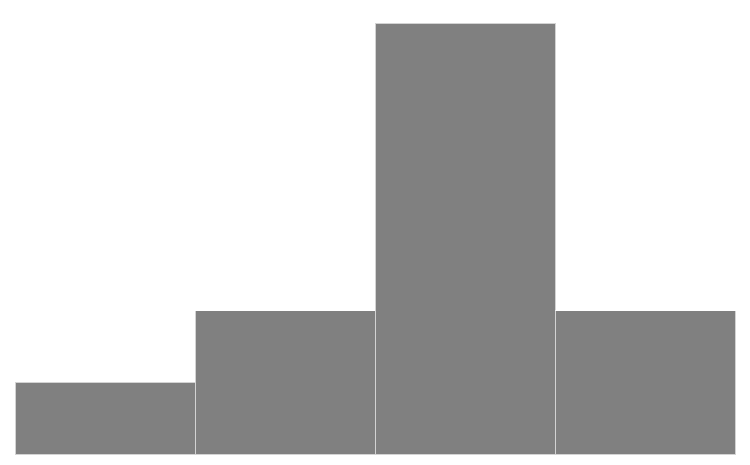}{To what extent it is true that most product/service features are invented rather than discovered?}.

The four bars denote the distribution of respondent agreement with a statement on a Likert scale (``not at all'', ``a little'', ``somewhat'', and ``very much'). The above example shows that most respondents ``somewhat'' agree with a statement (shown after the histogram), and the distribution of responses is skewed towards agreeing with the statement.








 \subsection{Team process area}
The team process area concerns start-up teams with respect to 
formation of a team, individual skills, attitudes, capabilities, and coordination between 
individuals. 

\subsubsection{Goals}

Establishing a team posing sufficient skills and expertise is one of the first goals of a start-up (G1). Earlier studies suggest that the team is the catalyst for product development in start-ups~\cite{Unterkalmsteiner}. Team characteristics such as cohesion, coordination, leadership, and learning are recognized as essential for software project success~\cite{dingsoyr2016team}.

Across our sample, the median team size is 4 - 8 people and 1 - 3 of them 
primarily work on software engineering. We observe a tendency of growing median 
team size over the start-up life-cycle, from 1 - 3 in the inception stage, to 4 - 8 
in the stabilization and growth stages, and to more than 20 in the maturity stage.


The responses suggest that to build a team, start-ups need to address 
communication issues, shortages of the domain and engineering expertise, 
commitment issues, and to create accountability. Statistical analysis on 
accountability in the teams shows an association ($Cramer's V = 0.334, p < 0.05$) between 
closed companies and a lack of accountability. Thus, forming a unit that poses 
relevant and complementary expertise, and that can work together efficiently 
with little overhead is an essential early goal in start-ups, see G1 in Fig.~\ref{fig_tpa-overview}.

By comparing responses from start-ups at different life-cycle stages, we found 
that establishing a team is a concern in the early stages of a start-up. We 
observe that start-ups at the inception and stabilization stages reflect on issues 
associated with creating a team, see G1 and C2 in Fig.~\ref{fig_tpa-overview}. However, start-ups in the maturity 
stage reflect on challenges related to managing a large team~(C1).

\subsubsection{Challenges}

Looking into respondent reflections on their team formation, we found that the majority, 61 out of 84, 72\%, reports some team-related challenges. The 
challenges concern team formation, management, expertise, leadership, and coordination, see C1 - C3. 




Start-ups at all life-cycle stages report shortages in engineering skills and domain expertise. The most severe shortages are reported by start-ups at the inception stage where only 3 out of 15, 20\%, respondents, estimate their engineering, and 1 out of 15, 6\%, rate their domain knowledge as sufficient.
However, we observe a tendency of estimates improving over life-cycle stages. A potential explanation could be survivor bias and start-ups who 
managed to acquire the necessary competencies were able to advance~\cite{shermer2014survivor}. We also observe a tendency that less skilled teams spend more time in the inception stage and often fail to release the product altogether. As one start-up reflected on their lessons learned from their team formation: 

\myquote{We should have looked for more developers with experience in delivering products fast, and do not rely on corporate ``experts'' that deliver anything but completed software.}

Statistical analysis shows a strong association between domain knowledge and 
engineering skills in the team ($Cramer's~V = 0.513, p < 0.05$). Teams with less domain 
knowledge also lack engineering knowledge, and teams with adequate domain 
knowledge are likely to have sufficient engineering skills. The level of 
expertise is also associated with the state of the start-up ($Cramer's V =
0.316, p < 0.05$). Operational start-ups estimate their skills and knowledge 
significantly higher than paused or closed companies. 
The level of team skills and experience is reported as one of the key success factors in software projects~\cite{Chow2008b}, and new business creation\cite{politis2008does}.
Thus, there could be a causal relationship between the level of team skills and start-up outcomes.

From the 41 start-ups in growth or maturity stages, 80\%, or 33 companies, reflect 
on the need for specialist skills and the challenge to acquire them (C3). For 
example, finding employees with domain-specific experience, or engineers with 
specific technical skills. Often start-ups reflect, that it would have been 
beneficial to acquire such expert skills earlier. However, the responses do not reveal the 
cause for the difficulty.

The shortage of experts with domain-specific experience could be explained by short supply of experts in
a narrow and potentially new area. Such explanation highlights the importance of education and training in start-ups. However, an alternative explanation could be that experts choose not to work with start-ups due to, for example, an uncertain future of the company.

By analyzing the responses on the team's attitude towards good engineering 
practices and their engineering skills we found that less skilled teams are 
less predisposed towards following good engineering practices such as avoiding code 
smells ($Cramer's~V = 0.342, p < 0.05$) or thorough testing of their product ($Cramer's 
V = 0.366, p < 0.05$). Further analysis reveals that the attitude towards following good 
engineering practices is associated with start-up outcomes. Operational 
start-ups recognize benefits from following good engineering practices, such as 
maintaining good software architecture ($Cramer's~V = 0.400, p < 0.05$) and avoiding code 
smells ($Cramer's~V = 0.320, p < 0.05$). Paused start-ups report seeing fewer benefits 
from good engineering practices. This finding suggests an association between 
start-up outcomes and attitudes towards utilizing good engineering practices.

Interestingly, respondents with more individual experience estimate their team 
attitudes towards good engineering practices and quality of engineering work 
significantly worse than less experienced respondents ($Cramer's~V = 0.384, p < 0.05$). 
This finding could be interpreted as follows: a) inexperienced engineers overestimate the quality of their and their teams' work, or b) experienced engineers underestimate their work. Kruger and Dunning~\cite{Kruger1999} have studied cognitive biases in estimating own abilities. Their findings suggest that low-ability engineers cannot objectively evaluate their actual competence or incompetence, and are likely to overestimate their abilities. 

We have also found that early start-ups do not implement any objective metrics to assess their team performance, see section~\ref{sec_results_pm}. Therefore, implementation, evaluation, and improvement of engineering practices depend on competence and gut-feeling of early start-up engineers.

Individual skills, competencies and teamwork capabilities have been recognized 
as essential success factors in software engineering 
projects~\cite{Wohlin2015,Society,Chow2008b}. Earlier studies suggest that 
start-ups rely on implicit knowledge, and have very little, if 
any, organizational capital~\cite{Unterkalmsteiner, seppanen2017little}. Thus, establishing a small 
and efficient team is essential to compensate for the lack of organization. Our results show that early team issues could be a reason for stalling product engineering and collapse of a company even 
before the product is launched to market.

Another challenge reported by 67\%, 56 of 84 companies, is to engage the team 
and coordinate the product work. The difficulty stems from team members 
having other priorities outside the start-up in combination with a bloated, 
poorly organized, and distributed team. The reported issues in such teams are poor 
communication, lack of clear responsibilities and unbalanced skill set with 
further effects on productivity, degrading motivation, and poor execution 
of multiple parallel activities, e.g., product engineering work and marketing, among other challenges.

From all team related concerns we distill 2 main challenges. The first challenge pertains team building and comprises of a lack of team expertise, engagement, and coordination at the inception and stabilization stages, see C2. The second challenge, see C1, is to manage a large and potentially distributed team at the maturity phase. Respondents mention difficulties to coordinate and maintain efficient teamwork across multiple teams and time-zone. Such findings suggests that start-up principals need to recognize and adapt for different teamwork challenges as the company moves forward.

Part of the challenge to initiate teamwork is the absence of leadership to 
establish an engineering team and to drive the product engineering work. As 
stated by one start-up: \myquote{All of the founders had other occupations. 3 
professors (2 of them located in the USA), and one business owner located in 
Turkey. There was a lack of communication. The developers were hired part-time. 
There was no-one for whom this start-up was their primary occupation. Maybe 
hiring a manager would have helped.}

Looking more into leadership we found 4 cases explicitly pointing out the 
need for technical leadership. A technical leader, or CTO, is needed to set up 
an engineering team and to lead product engineering work. As one respondent, 
employed by a start-up, stated: \myquote{I would fire the current CTO and hire 
a new CTO, who is more technically 
sound. In the startup, the most prominent thing is that the CTO 
should be technically sound. Otherwise, it is hard to drive the 
product forward and to motivate the development team.}

A potential pitfall is to select the technical product leaders based their executive powers and not professional competences. This could be the case when, for example, a less competent founder refuses to give away the CTO role to a more suited employee~\cite{Crowne2002}.

An association between good teamwork practices and software project cost and quality is studied
in the context of established companies~\cite{krishnan1998role}. Higher team capabilities regarding technical skills and domain 
knowledge, are strongly associated with a lower number of discovered defects and 
lower software maintenance costs. Moreover, commitment to a shared goal and 
internal team communication mechanisms are essential for project success. Our results
indicate that such findings are relevant in the start-up context as well.

We looked into how the respondents' relationship with the start-up affects their 
responses. We found that founders of start-ups are significantly more 
optimistic about the quality of planning ($Cramer's~V=0.381, p < 0.05$), and quality of 
product engineering ($Cramer's~V = 0.385, p < 0.05$), compared to hired engineers and 
external contractors. Such results suggests a potential fault-line in the teams between 
founders and employees in terms of how they perceive the engineering context. 

As shown by Chow~\cite{Chow2008b}, joint decision making and knowledge sharing, critical to project success,
depend on efficient communication. However, fault-lines splitting a team into two or 
more sub-groups, result in impaired communication and further adverse effects from stemming from communication issues.
Causes and effects of team fault-lines are observed and studied in the context of globally distributed teams, for example Gopal, et al.~\cite{gopal2011coordination} and Staats et al.~\cite{staats2012team}. 

Earlier studies suggest that start-ups have small, flat and empowered 
teams. Empowerment of individuals is supposed to reduce the need 
for bureaucracy and improve flexibility~\cite{Paternoster2014}. 
However, our results suggest that even though 
teams are small, there still exists a communication gap between founders and 
employees. An explanation could be that principal decisions from founders could be ill-communicated, thus 
perceived by employees as unjustified, and degrading motivation and trust in a team~\cite{boies2015communication}.

\subsubsection{Practices}

We looked at practitioner responses to identify how team formation challenges 
are addressed in their start-ups. We recognize two general scenarios how 
start-up teams are formed (P1). One scenario is creating a new team from people 
without previous joint experience. The second scenario is that a team 
originates from a former organization and already has some teamwork experience.
Not every start-up could have the opportunity to reuse an existing team, thus 
we consider these both strategies as variations of the same practice to establish a team. 

The responses suggest that newly created teams start with a few founders and new 
people are added when there is a need for additional skills or human resources. Such 
organizations have yet to establish teamwork practices, acquire domain knowledge, and 
vet their engineering capabilities. Thus, new teams are prone to teamwork 
issues, shortages of skills and expertise. 

Teams originating from earlier projects are slightly larger and have more diverse 
competencies (compared to new teams), shared history concerning established 
ways of working, roles, responsibilities, and had already ironed out initial 
team formation issues. While we do not know the exact relationships between team 
members in all cases, 9 respondents mentioned that their teams have been 
working on earlier projects suggesting that they have shared experience before
the current project. An example of this scenario would be a small consultancy 
company, i.e., offering customized services, that identifies an opportunity 
to develop a product for mass-market. They keep the consultancy business going to 
support the start-up endeavor, eventually aiming to become a product 
organization.

Making use of an existing team helps to alleviate initial team formation 
challenges, see C2 in Fig.~\ref{fig_tpa-overview}, and minimize the risk 
of the team breaking apart. Moreover, an existing team likely has experience in 
relevant product engineering technologies, markets, and the product domain. 
Therefore, such teams have an advantage over recently formed teams with no 
shared background and experience. Similar results, pointing out that start-up 
founders' earlier experience shapes their skills and attitudes helping to cope 
with uncertainty are reported by Politis~\cite{politis2008does}. However our 
results show that skills and expertise of the team as a whole plays an 
important role.

Start-ups report different tactics to address the lack of engineering 
competences. As an alternative to establishing an own engineering team, 3 
start-ups mention outsourcing product development work to another company (P3). 
The motivation for outsourcing is to quickly build the first version of the 
product without the effort of creating an own team. However, outsourcing the 
engineering work comes with challenges to negotiate requirements and 
communicate efficiently. As one respondent stated: \myquote{We started the 
development offshore with an external company, now 
development is in-house. With fewer people, we have the same cost, 
but we at least tripled the productivity.}

Start-ups at all life-cycle stages mention the use of external consultants to help with 
specialist tasks (P2) in addition to their engineering team. For example, 6 start-ups 
mention that they have used external user interface specialists to help with 
product design. Some mention using external developers for mobile application 
development, security, and optimization related tasks.

\subsubsection{Lessons learned}




Our results support earlier findings that team is the catalyst for product engineering~\cite{Unterkalmsteiner}. Team issues could hinder start-up potential to advance through the life-cycle stages.

For start-ups, our findings present several implications: 
\begin{enumerate}
  \item Team formation is an essential early activity. To attain a highly
    performing team, a team building program must be implemented, focusing on establishing respect for everyone in the organization, identify and communicate individual performance standards, develop ways of efficient communication, identify clear individual and group goals, reward teamwork, and team-building efforts, and encourage loyalty to the team~\cite{tippett1995team}. \\
  Suggested reading: Bubshait et al.~\cite{bubshait1999team} presents critical concepts influencing team performance and lists building blocks for establishing highly performing teams.

  \item Engineering and domain expertise are important for efficient teamwork. 
  Engineering expertise is essential to build the product fast. However, domain 
  understanding helps to identify and interpret software 
  requirements~\cite{hadar2014role}. Start-up teams are often formed by 
  inexperienced people or work in new domains, thus knowledge sharing and 
  mutual learning is essential. \\
  Suggested reading: Eppler et al.~\cite{eppler2000managing} presents 
  processes, tools, and factors for enabling team knowledge management. Cockburn 
  and Highsmith~\cite{cockburn2001agile} explore the role of individual 
  competences in agile team performance. 

\end{enumerate}

\subsection{Requirements engineering process area}

The requirements engineering process area concerns the elicitation, analysis,
validation, documentation and scoping of software requirements. The
identification and validation of a relevant product idea, that is requirements identification, validation, and scoping, are some of the most important activities in start-ups~\cite{Blank2013}.

\subsubsection{Goals}


One of the first steps in any software project is to identify needs and
constraints placed on a software product~\cite{Society}. 
Respondents from
start-ups at the inception stage agree that product features are to a large extent invented, and are based on founders experience, and understanding about the domain~\sparkline{figures/sparklines/spark_0_DA.pdf}{To what extent it is true that most product/service features are invented rather than discovered?}.

As a consequence of invented requirements, one of the first objectives in a start-up is to break down invented ideas into software requirements and to validate these requirements, optimally with inputs from target customers (G2). As one practitioner stated: 
\myquote{The real purpose of requirements elicitation is to invalidate requirement ideas as quickly as possible without unnecessary effort on meta documentation or implementation.}

Responses show that as soon as the product is launched to market, start-ups put 
more emphasis on using their customers as requirement sources. Requirements 
invention becomes less common. Responses, to what extent product requirements are invented, shifts towards disagreement when start-ups mature.~\sparkline{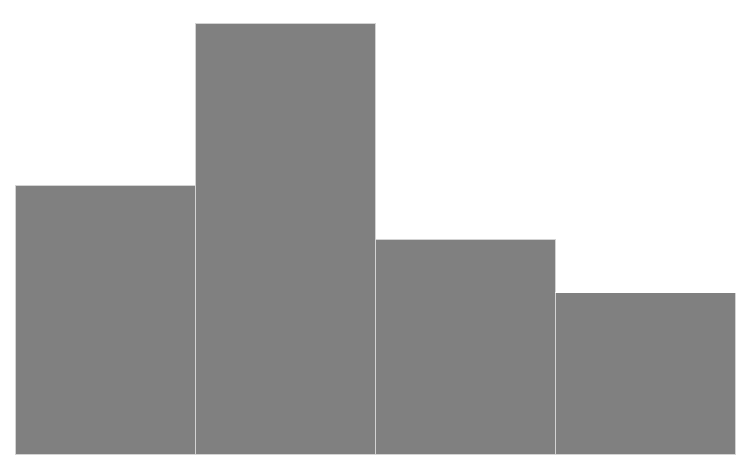}{To what extent it is true that most product/service features are invented rather than discovered?}. Thus, the goal to quickly invalidate own ideas is most relevant at the inception stage and before start-ups have established a feedback loop and could use customer input for feature ideas, see C4 and P5 in Fig.~\ref{fig_tpa-overview}.

Requirements validation is mainly about what functionality and quality to offer. However, it is essential to differentiate between the validation of the idea itself (requirements) and how it is provided (design)~\cite{vanLamsweerde2003}. The differentiation is significant. A great idea can be received poorly if the design of it, regarding how the solution is offered, is bad.

Results from established companies working on market-driven products are similar. Mature companies, alike start-ups invent or discover requirements indirectly through analysis and observations. Established companies face similar challenges to validate requirements before a product is launched. This is compensated by internal requirements analysis and frequent releases~\cite{Dahlstedt,Alves2006}.

To achieve the goal of the inception stage and to release the first version of
the product, see Fig.~\ref{fig_states}, start-ups must determine the scope of
the minimum viable product (MVP). The MVP is a trade-off between features,
quality, time, and cost~\cite{Junk2000} used to gauge market interest in the product and to establish an early customer base justifying further investments in the product.

When asked how the MVP was scoped and what quality attributes were considered necessary, the majority of respondents, 56 out of 84, 67\%, pointed out that the priority is to maximize customer value through functionality, usability, and user experience, while keeping engineering effort minimal, see P6. Practice of scoping the MVP is related to the goal of balancing customer value with minimal development effort (G3). As one respondent reflects the MVP should be scoped by the current needs, and not by wishful thinking:

\myquote{Any requirement that is essential for validating the growth or value
  hypotheses is priority. Anything that is to support the business when the user base is larger than 100 users is not a priority. Scoping is an exercise of un-prioritizing features from being added to the MVP.}


Responses on release scoping goals from start-ups in stabilization, growth and
maturity stages suggest a shift in scoping goals compared to the inception
stage. The release scoping goals of inception and stabilization stages are to
maximize customer value and to validate the product idea, however later goals
shift to supporting business goals, such as monetization and growth (G4). We
present related results on value focus and product quality goals in Sections~\ref{sec_results_value} and~\ref{sec_results_quality}.



\subsubsection{Challenges}

Comparing responses from active and closed start-ups at stabilization and growth stages we observe that internal sources, such as brainstorming and invention of requirements, are the most popular requirement sources and used by 94\% of active, and 71\% of closed start-ups. 

Other requirement sources are similar products, used by 83\% active and by 43\% closed start-ups, and market trends, used by 57\% active and by 29\% closed start-ups. However, only 43\% of closed start-ups have used input from potential and existing customers compared to 91\% of active start-ups. This difference is statistically significant ($Cramer's~V = 0.463, p < 0.05$). 

Looking at the free text responses for an explanation, we found that all closed start-ups, and 10 out of 35, 29\% of the active start-ups had difficulties establishing contact with their potential customers and involving them in the product work (C4, P5). The common shortcomings are involving customers too late in product work, for example, only after release to market, and sampling of potential customers for requirements elicitation and validation. As one respondent reflected: 

\myquote{We informally asked our friends and family about existing solutions, then we brainstormed with the information collected about how to design the product and what features it should have. We never used a formal method to elicit requirements, we just informally decided the features our product should have.}

The importance of stakeholder involvement in new product development as a success factor has been pointed out by nearly every study on requirements engineering, for example, Chow~\cite{Chow2008a}, Karlsson~\cite{Karlsson2007}, Hoyer~\cite{hoyer2010consumer}, and Blank~\cite{Blank2013b}. However, as our results show, finding the right sample of customers and convincing them to invest time in collaboration is challenging. 

Establishing relationships with potential customers is a goal of sales activities as well. A recent study on new product development from the sales perspective highlights the overlap between requirements engineering and sales processes, pointing out that getting to know customers is essential for both technical and commercial outcomes of the project. Involving both engineering and sales roles in establishing customer contacts helps to ensure continuity between requirements engineering and commercial relationships~\cite{la2016customer}.

Some, 7 of 24, 30\%, start-ups at the stabilization stage reflect that they had believed that adding more features to the product would improve their chances of success. That constitutes feature creep, see C5, and stems from difficulties to elicit useful feedback from customers, and generalizing customer specific requirements. Start-ups report that feature creep drained financial resources and added extra complexity to the product. As one company reflected on their lessons learned: 

\myquote{We should have done usability tests with real customers with a prototype of the product and iterate faster based on user feedback and actual usage, not just guessed what customers might want.}

Feature creep is reported as a general challenge in developing new software products. It is caused by adding new requirements late in the product development without appropriate analysis. The new requirements could stem from external stakeholders, thus appear very appealing to include in a release, or could be discovered and self-approved by the team during the development process. Either way, a company should analyze the impact and assure that higher business goals are not compromised~\cite{elliott2007anything}.


\subsubsection{Practices}


Responses suggest that start-ups use a mix of requirements sources. Across all cases, internal sources such as requirements invention and brainstorming are the most reported by 76 out of 84, 90\%, start-ups, followed by potential and existing customers (66 cases, 79\%), and analysis of similar products (59 cases, 70\%). Market trends and business goals are less utilized, standards, laws, and regulations are used by start-ups in regulated domains, such as medicine.

Respondents suggest that they have used their previous experience in the domain, both professional and from using similar products, to identify ``obvious'' requirements and requirements sources. However, customers are the most valuable source of requirements, as stated by one respondent: 
\myquote{The most important source is existing and potential customers 
where we have continuous dialog in place. This is somewhat 
self evident. Understanding of competition, 
business models and regulations is secondary to that.
}

\textit{Elicitation, validation: }
Start-ups report using observation (64\%) and interviews (61\%) to elicit requirements from customers. On-site customer and surveys are less used, 32\% and 46\% respectively. Prototypes and mock-ups are used by 61\% of respondents to support brainstorming and to elicit feedback in customer interviews. 

Responses suggest that elicitation is triggered by internal ideas that are further elaborated and iterated with input from customers. As one company reflected: 
\myquote{Elicitation is a cyclical process of getting info from customers/products, analyzing and then brainstorming to feed into prototypes and mock-ups. Early in the project, this was at a very high level. Later when we concentrated development on specific features, the steps were repeated in a more detailed manner.}

Requirements elicitation overlaps with requirements validation (P4). Or, as one
respondent put it: ``the elicitation techniques are used to IN-validate
requirements/ideas/features''. The most frequently reported validation technique
is internal reviews used by 55\%, 46 start-ups, followed by prototype
demonstrations to customers reported by 49\%, 41 start-ups. A/B tests to measure
customer reaction on new features are used by 25\%, 21 start-ups. Use of A/B
testing increases over the start-up life-cycle from 6\% at inception stage to
34\% at the maturity stage, potentially because A/B tests and other data-driven methods require a significant number of customers interacting with the product and such numbers may not be available at early stages~\cite{olsson2014opinions}.

By looking into associations between difficulties in requirements elicitation and demographical information we found that younger respondents, 25 - 34 years old, report more difficulties in collecting and prioritizing requirements than older, 35 - 44 years old, respondents ($Cramer's~V = 0.599, p < 0.05$). An explanation for this finding could be that older age is associated with a broader network of personal contacts and more extensive domain experience supporting identification of relevant ideas for product features. 

Similar results are reported by Azoulay et al.~\cite{hbr_age} suggesting that with older age comes more experience, business acumen, broader social network, and greater access to financial resources, thus older founders are more likely to succeed commercially. 

Start-ups in all life-cycle stages report that requirements are changing and they have an informal process to manage changes (P7). The most common source of changes is input from customers. Responses suggest that start-ups aim to work in short iterations and frequently re-prioritize their backlogs. Thus, requirements changes do not have any significant adverse effects.
As one respondent described their change management process: 
\myquote{Yes, requirements change! We happily abandon the obsolete requirements 
and remove any code from the product if there is any. New requirements 
are re-prioritized with a goal to validate our growth and value hypotheses.}


From our sample, only 7\%, 6 start-ups,  have explicitly stated that they do not
document their requirements in any way. The most used specification format is
informal notes and drawings, reported by 56\%, 47 respondents, others report
more formal techniques, such as the use of templates and formal specifications, for
documenting requirements. Most commonly, start-ups document requirements  on a
feature level (43\%, 35 cases), and a function/action level (21\%, 18 cases). Requirements are written down as ideas which are later elaborated (P7). As one respondent reflected on their practice:

\myquote{We used an on-line tool (Trello) to write all the features we wanted to implement and, at the same time, to organize their development. Trello cards served as the requirements to be implemented. There were no steps in documenting requirements per-see, we just talked informally about what to implement and then wrote it down in order to not forget it.}

With further statistical analysis, we found that the understandability of
requirements is associated with the state of a company ($Cramer's~V = 0.319, p < 0.05$).
Significantly more closed companies, compared to active ones, report that even though requirements were written down, they were difficult to understand and use in practice. Teams with insufficient domain knowledge are also more likely not to document their requirements. However, teams with adequate domain knowledge are more likely to use templates for documenting their requirements ($Cramer's~V = 0.345, p < 0.05$). These findings suggest that more rigorous requirements documentation could be a way to acquire, document and distribute critical domain knowledge in the team. Alternatively, teams with better skills and domain knowledge see the upside of documenting requirements. Either way, we observe an association between understandability of requirements, improved domain knowledge, and progression of a start-up.

Requirements documentation enables to create a plan, outlining what features to
implement when, i.e., develop a product road-map.  The responses suggest that 46 start-ups, 55\%, have such a road-map. From start-ups at inception,
stabilization and growth stages, about half, 40 - 50\% of the star-ups, report having a road-map. However, product road-maps are used by nearly all, 87\%, of mature start-ups. Between closed and active start-ups, 67\% of operational, 40 cases, have road-maps. However, only 2 or 18\% of closed start-ups had road-maps.
Respondents reflections suggest that road-maps are used as planning documents internally and often synced with primary stakeholders. Parts of a road-map that concerns near future are more detailed, however long-term plans are described at a higher, milestone level.

As suggested by the start-up life-cycle model, the first significant milestone for start-ups is to release the minimum viable product, see Fig~\ref{fig_states}.  
Responses suggest many strategies for scoping the MVP (P6), such as using their domain knowledge, input from potential customers and partners, soft launch by releasing the product to one customer at the time, time-boxing, and ``gut-feeling''. With the MVP, start-ups aim to deliver the essential features in a shortest time possible. As one practitioner described the MVP scoping process: 
\myquote{We carefully removed any feature or function not essential to testing growth and value hypotheses. Of which two, the value hypotheses is prioritized.}

Value is specified as the primary prioritization goal by 85\%, 71 of 84, of
respondents. Implementation time is the secondary goal reported by 38\%, 32
cases. Requirements prioritization is done by consulting customers and other
stakeholders. We observe that at the inception and stabilization stages,
start-ups consider customer needs, while at the growth and maturity stages, business requirements, such as a need for revenue and growth, are discussed as well. As one practitioner described their prioritization process: 
\myquote{To prioritize we use this question: how much money or new customers we will have if we implement the new feature?}



In their reflections, nearly all respondents from both active and closed start-ups, suggest that they should have spent more time with customers to understand and analyze their needs better. 
As one respondent describes their lessons learned:
\myquote{In the start we let innovations lead our goals too much. Then we got
  our first customers and did not listen to them much. We did not track in a
  enough detailed way what the customer is exactly doing with our product. We
  let just one dedicated person to be in contact with certain customers and did
  not share details in the team. Even a small company needs to setup a program to make interaction with customers transparent and actionable.}


Comparing our results on requirements engineering in start-ups with results from established market-driven companies we observe many similarities. For example, in both contexts requirements are primarily invented, used practices are light-weight and informal, and focused around quick releases to elicit customer feedback~\cite{Dahlstedt,Alves2006}. However, we observe a difference in prioritization practices. Established companies rely more on effort estimates in prioritizing requirements, while start-ups use value as the primary prioritization target~\cite{dos2016requirements}.

The discrepancy in prioritization goals could be explained by a lack of unified and quantifiable view on
value. Thus, established companies opt for scalable and straightforward
prioritization criteria~\cite{Khurum2012}. Moreover, established companies are
likely to operate within a set budget and schedule constraints further motivating
the need to adhere to effort estimates. However, start-ups are more customer-centric, flexible, and work on a smaller number of features, thus can use value as prioritization target~\cite{Giardino2014}.

\subsubsection{Lessons learned}
Our results support earlier findings that requirements engineering is one of the key engineering activities in start-ups~\cite{Melegati2016}. Moreover, our results show an association between requirements engineering practices and state of a start-up. We present the following implications for practitioners: 

\begin{enumerate}
  \item Starting to collect input from potential customers as early as possible is the key to identifying the most relevant requirements. By involving
    customers in new product engineering, companies can achieve a higher degree of efficiency, ensure product fit with customer needs, and achieve higher customer engagement and satisfaction~\cite{hoyer2010consumer}. Moreover, early customer relationships are a basis for sales activities when the product is launched.

  Suggested reading: Cui et al.~\cite{cui2016utilizing} compiles earlier work from marketing and innovation literature and presents practical guidelines on how to involve customers and use their knowledge in developing innovative products.\\
  Hoyer et al.~\cite{hoyer2010consumer} present a framework for value co-creation in product development comprising of motivators, outcomes, and potential impediments. 

  \item Feature creep can be managed by requirements analysis focusing on how many customers will find a feature useful. Only features that concern the majority of the customers should be included in the roadmap. 

  Suggested reading: Elliot~\cite{elliott2007anything} proposes to use the Pareto principle in deciding whether a feature is part of the core product or is customer specific. He argues that about 20\% of features are used by 80\% of customers, and the key to avoiding scope creep is to pinpoint the 20\%. Also, the paper lists several strategies for handling feature creep.

  \item Writing down requirement ideas, their source and rationale can help to acquire, maintain and distribute domain knowledge in the team. For example, even a basic requirements specification helps to establish a common vocabulary and avoid delays and time-consuming interactions caused by confusion and misunderstandings between stakeholders~\cite{buchman2009barriers}.

  Suggested reading: Hadar et al.~\cite{hadar2014role} explore the role of domain knowledge in requirements elicitation. They analyze both positive and negative effects of prior domain knowledge in elicitation interviews. Domain knowledge helps to ask more focused questions, provides a common language, and saves time in learning the basics. 

  \item Release scoping, especially scoping of the minimum viable product, should be done carefully and optimally with clear goals and input from all stakeholders. Our results from start-ups are similar to results from established companies suggesting that facing uncertainty companies are likely to overscope their product releases~\cite{bjarnason2010overscoping}.
  Suggested reading: 

  Suggested reading: Bjarnason et al.~\cite{bjarnason2010overscoping} presents a root-cause analysis and effects of release overscoping in software projects.

\end{enumerate}

\subsection{Value focus}\label{sec_results_value}

The software value concept characterizes the broader aims of a company and aligns all activities in an organization towards defined value goals~\cite{Boehm2003}. Respondent responses often mention value as a criterion for prioritizing requirements and scoping product releases. However, a value is a vague term and could mean different things to different stakeholders. To explore how start-ups interpret the phrase we map their definitions of value to software value aspect taxonomy~\cite{Khurum2012}. We summarize these different views as engineering goals.


\subsubsection{Goals}

Across the sample, the dominant view on value is the customer perspective, reported by 48\% or 40 start-ups. Respondents define customer value as perceived benefits, regarding functionality, user experience, and hedonistic value, derived from the product, and potential for the company to capitalize this value, see G5-G8. As one respondent phrased it:

\myquote{With value we understand the benefit for a customer to use our product.
  Higher value strengthens our position in the market compared to competitors and helps to increase the price of the product.}

The second most reported interpretation of value stems from the internal business perspective, i.e., how the company estimates product value from their own, internal perspective (23\% or 20 start-ups). Respondents define this perspective as both market potential, e.g., how many customers could benefit from a particular feature (G6), and differentiation value, e.g., how a feature will help to stand out in the market (G8). As one respondent put it:

\myquote{Value is something that can be used to market/distinguish the product from competition.}

Financial value concerning revenue is the third most reported interpretation of value, by 11\% or 9 respondents (G7). However financial value is often defined in combination with other value perspectives, such as customer or internal value. As one respondent described their multi-faceted view on value:

\myquote{Value is what could give more revenue to the company and make the product easier to use for users and motivate them to purchase, so that we could have more transactions each day.}

We compare value definitions between start-ups at different life-cycle stages and observe several tendencies. 
Customer value pertaining perceived benefits of the product of is the dominant value perspective in all life-cycle stages,
reported by 23 - 46\% of respondents. 
Internal value capturing product's market potential value, for example, ability to serve more customers efficiently and to access broader markets, is reported by 20\% of respondents at inception and growth stages, 12\% at the growth stage, and not reported at all by start-ups at the maturity stage. 
Financial value of the product is not reported at all by companies at inception stage, however it grows over start-up life-cycle and peaks at the maturity stage where 20\% of start-ups have reported it.
Internal, differentiation value are more reported by start-ups at inception and maturity stages, and appear less relevant at stabilization and growth stages.


A similar analysis conducted in established product companies shows somewhat
different results~\cite{alahyari2017study}. While the ranking of value is
similar, customer value being the most important, followed by internal value and
financial aspects, we observe differences in the definition of these values. For
example, established companies interpret customer value as delivery time and
perceived quality. However, none of the start-ups in our sample have mentioned
time-to-market or product quality in their value definitions. The internal value in
established companies is understood as internal product quality, technical debt,
supporting tools and processes. Start-ups, in turn, focus on market potential and differentiation value.

Differences in the studied samples can explain these discrepancies in value focus. Established companies could see more value in consistently delivering quality features to existing customers and keeping technical debt low. However, start-ups aim to identify an unmet customer need in a high-potential market~\cite{Sutton2000, Blank2011}.

\subsubsection{Lessons learned}\label{sec_value_learned}

Our results show shifting views on value definition among start-ups in different life-cycle stages. However, there is a shortage of similar results (except for Alahyari et al.~\cite{alahyari2017study}) for comparison and deeper analysis. Nevertheless, we present the following implications for practitioners and gaps for further investigation:

\begin{itemize}

  \item Understanding of value focus can help practitioners to the scope and align their engineering and business activities to maximize specific types of value. Alignment of activities helps to ensure that technology actually
    supports specific business goals and deliver the intended value to stakeholders~\cite{Carlson2006}.

  \item Our results show that at least two value perspectives are relevant at any life-cycle stage. The value focus shifts from customer value and internal market value at inception stage to financial and differentiation value at the maturity stage, see Fig.~\ref{fig_tpa-overview}. Understanding of the multi-faceted nature of value can help to facilitate communication between stakeholders~\cite{Khurum2012}. 

  Suggested reading: Khurum et al.~\cite{Khurum2012} present a taxonomy of software value aspects establishing a common vocabulary and understanding on different value aspects.

  Carlson and Wilmot~\cite{Carlson2006} present a practical guide on how to identify, use and develop an understanding of the value and use value to align organizational efforts in building innovative products.

\end{itemize}

\subsection{Quality goals and testing process area}\label{sec_results_quality}

This process area concerns product quality goals and practices to attain these goals. Product quality is a mix of functionalities, non-functional attributes, and broader constraints determining commercial success of a product (e.g., cost of product development vs. returns from marketing the product). We aim to explore what aspects of software quality are considered significant by practitioners and what practices are used to attain such aims.

\subsubsection{Goals}



Respondent answers suggest that product functionality is the most common quality goal (G9) across all life-cycle stages, reported by 32\%, 27 respondents. Time-to-market is the second most common objective (G11), indicated by 15\%, 13 start-ups. Even though time-to-market is not a product quality goal intrinsically, it has a profound influence on the product decisions~\cite{Unterkalmsteiner,Carmel1994c}. The frequency of other quality goals varies across life-cycle stages.

Maintainability is frequently reported quality goal in stabilization, growth and maturity stages (G10). Portability is reported only by start-ups at the maturity stage (G12). Reliability is mentioned by few cases in stabilization and maturity stages. Shifting quality goals indicate suggest changing priorities and adds support for start-up life-cycle model, see Section~\ref{sec_model}.

Looking further into how the required level of product quality was determined we
found that start-ups reflect on a ``good-enough'' level of their quality goals
(G13). This goal is related to scoping of the MVP (P6) and focus on
non-functional features of the product. Wrongly estimating the required quality level (G13, P8) could lead to poor market reception due to less than acceptable quality, or waste by providing excessive quality~\cite{Regnell2008}.

\subsubsection{Practices}

Start-ups report using different strategies, based on in-house expertise, user feedback, and iterative development (P8) to set their quality targets.

One strategy is not to set any specific quality targets and iteratively identify, and improve relevant quality aspects. As stated by one company:
\myquote{We take a continuous iterative refinement approach rather than 
setting a fixed goal, so we periodically assess which areas are 
in need of improvement, evaluate and prioritize the options}

As an alternative, one start-up at the inception stage reported aiming for the simplest solution that can be improved, if needed:
\myquote{We look for what is the simplest, regarding size and complexity, solution to this problem. Will that solution be able to support a couple of hundred users? How many different ways can the solution be improved to 
support more users?} 

Regnell et al.~\cite{Regnell2008} describe a quality requirements roadmapping
model for determining the required quality level. The method proposes to
identify relevant quality metrics and defines useless, useful, competitive,
and excessive quality ranges for each. Then, it uses these ranges to compare the
own product with competition and to spot opportunities for improvement. Using
such a model in a start-up could help in determining essential qualities, a minimum
quality level for a product to be useful, and opportunities to differentiate
among similar products. This is aligned with the focus of internal value
start-ups typically have (see Section~\ref{sec_value_learned}).

Respondent answers suggest that informal manual testing is the most common practice for making sure that the product has an acceptable level of quality at all life-cycle stages, reported by 33\%, 28 start-ups (P9). However, the responses suggest that informal testing is gradually replaced with an organized QA process (P10) at the maturity stage. Alternative methods, often used in parallel, are exploratory testing, or scenario-based testing, reported by 21\% and 19\% start-ups respectively. 


Respondents estimates suggest that test case documentation varies from informal to systematic without any clear tendency~\sparkline{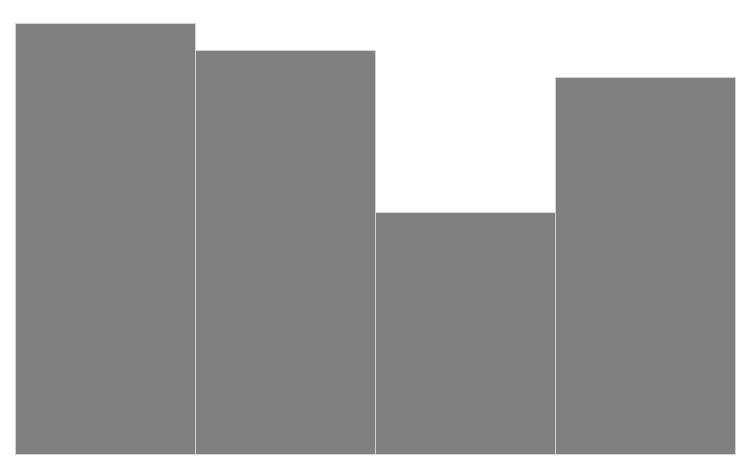}{To what extent it is true that test cases are not systematically documented?}. Estimates on the test case coverage are biased towards agreeing with less than complete coverage~\sparkline{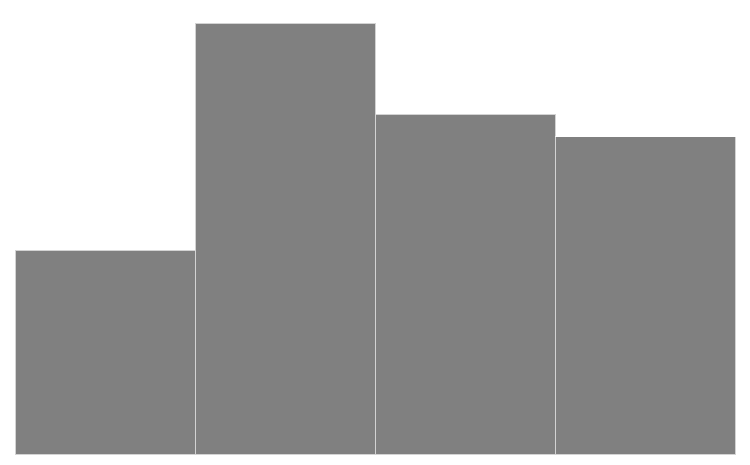}{To what extent it is true that test cases do not fully cover the product functionality?}

We observe little differences in testing practices between start-ups in
different life-cycle stages. However, start-ups at growth and maturity stages
reflect that more test automation, more skills regarding software testing, and
more systematic testing would have been helpful. An explanation for such results
could be that consequences of the informal testing surface only when a product gains
a user base. Two start-ups at the maturity stage reflect that a dedicated tester role, responsible for performing testing tasks, is needed (P10). As stated by one of the respondents:

\myquote{I think we need more structured testing, a manager of testing that coordinates the efforts, all being responsible [for testing the product] is NONE being responsible.}

Answers to questions on the use of automated testing show that a third, 26 out of 84, 31\%, of start-ups are attempting to implement automated testing, and only 17 start-ups, 20\%, have explicitly stated that no test automation is used.

Responses suggest that regression testing is an increasing concern over 
start-up life-cycle (C6). Start-ups at inception and maturity stages report that 
they spend substantial effort on manually testing the entire product \sparkline{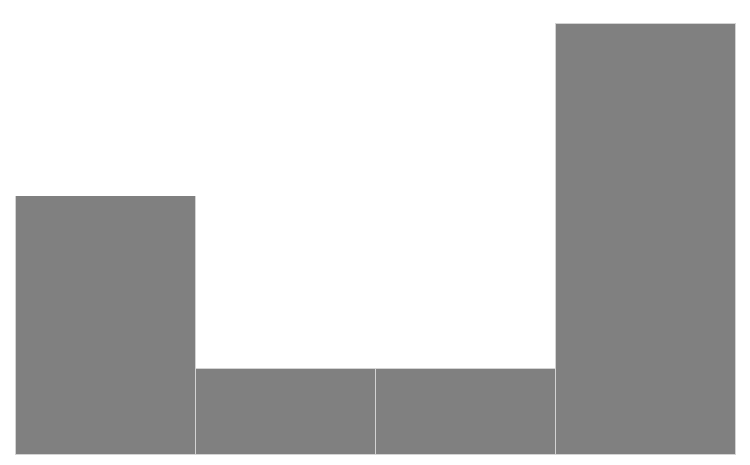}{To what extent it is true that manual testing of the entire product/service is required to make sure that a release is defect free?}. At the same time, respondents report that few defects slip through testing and are reported by customers~\sparkline{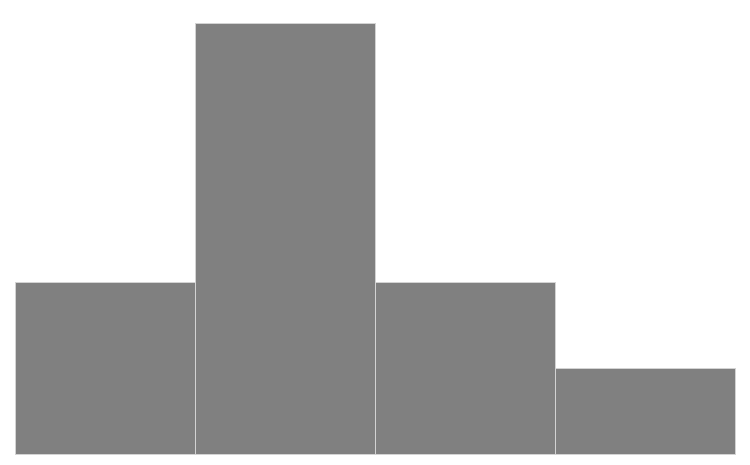}{To what extent it is true that customers often report defects that could have been captured earlier?}. Two respondents from start-ups at the maturity stage have stated explicitly that they are working to improve and automate their product testing.

Start-ups could benefit from more rigorous testing practices. Efficient software testing enables faster product releases, thus allowing the teams to reduce time-to-market and to iterate new features faster. More rapid release cycles contribute to faster requirements validation (G2, P5) and are known to improve customer satisfaction~\cite{chen2015continuous}.

Good testing practices and use of automated tests can support the on-boarding of new developers. Automated tests provide a safety net for inexperienced developers to discover any problems with their code quickly on their own. Automated test definitions serve as means of documentation to learn how different components of a product work~\cite{pham2017onboarding}. Therefore, having good testing practice and test automation have benefits beyond a defect-free software.

\subsubsection{Lessons learned}

Our results show that start-ups primarily focus on delivering relevant
functionality. Other quality aspects change as a start-up advances through the life-cycle. Software defects aren't a concern. However, the internal quality and testing practices are important to support the sustainable evolution of the product. We identify the following implications for practitioners:

\begin{itemize}
  \item Our results suggest that external quality isn't a concern in start-ups, potentially due to relatively small products and early adopters being more tolerant towards defects. That said, quality becomes a concern in growth and maturity stages when commercial success of a product depends on quality service. 

  \item We observe that maintainability is an increasing concern over a start-up's life-cycle. Maintainability helps a start-up to remain fast in launching new features and sets a foundation to enable portability of a product. 

  Suggested reading: Regnell et al~\cite{Regnell2008} propose a lightweight method for quality requirements roadmapping. The method helps to establish a frame of reference for assessing current product quality and spotting opportunities for improvement.

  \item Respondent answers suggest a lack of test automation and superficial testing practices. However, our earlier study on technical debt in start-ups could not find any significant association between testing debt and product quality or team performance issues~\cite{Klotins2018}. Nevertheless, good testing practices help to have faster product releases and speed up the on-boarding of new developers.

  Suggested reading: Collins et al.~\cite{collins2012software} present an experience report on test automation practices in an agile development context. They present several lessons learned from implementing test automation and what types of test automation bring the most benefit.

\end{itemize}

\subsection{Architecture and design process area}
The architecture and design process area concerns the internal product
structure, selection and use of components, construction technologies,
interfaces and other aspects supporting the construction of the product.

Earlier studies suggest that start-ups leverage on open-source components, third-party
services and cutting-edge technologies to construct their
products~\cite{Unterkalmsteiner}. We aim to explore how start-ups choose the
components and how they design their product architectures.

The visual appearance of graphical user interfaces influences the usability of a product and affects stakeholder perception about the product. Look and feel of product user interfaces is known to have an impact on project success~\cite{ralph2014dimensions}. We explore how start-ups design user interfaces for their products.




\subsubsection{Goals}

We inquired respondents on the use of cutting-edge technologies and found that start-ups aim to minimize risks from using immature technologies by using well-known, stable technologies in the product development (G14). However, 39\%, 33 of 84, of the respondents report using or experimenting with some cutting-edge components on the side. As one start-up reflected:

\myquote{We don't use unstable components or new 3rd party services without evaluating them thoroughly and running a small pilot project.}

For at least one start-up in our sample, poor choice of a technology platform
hindered quality, delayed product releases, and was the leading cause for
discontinuing the product. Therefore, selecting a technology stack is an
important goal early in the start-up life-cycle. 

Earlier studies have pointed out that start-ups leverage on cutting-edge technologies to develop innovative products and gain competitive advantages~\cite{Unterkalmsteiner2014,Sutton2000}. However, we could not find support for such results in our dataset. 

Looking into how organizations make technology decisions in other contexts, we found that performance regarding total time and resources efficiency, maintainability and reliability are the top criteria for software component selection in established companies. Similar results are reported by Petersen et al.~\cite{petersen2017choosing}, and Badampudi et al.\cite{badampudi2016software}. 
Start-up companies could be considering similar criteria and opting for more stable components to save development time and resources. Furthermore, start-ups could be leveraging on established, well-known technologies to create new and innovative products~\cite{maranville1992entrepreneurship}.

\subsubsection{Challenges}

Respondent estimates suggest that technical debt is prevalent in start-ups, especially late in the life-cycle, see C7.
Nearly all start-ups report some signs of technical debt, such as shortcomings
in knowledge distribution, code smells, suboptimal architecture decisions, and
lack of automated testing.  Testing debt, associated with lack of automated
regression testing and a need to manually re-test the entire product before releases, is the most common type of technical debt. Documentation, architecture debt, and code smells are also a concern, albeit to a lesser extent. 

In our earlier study with the same dataset, we found documentation, architecture debt, and code smells to be associated with impaired team productivity and product quality. However, we could not find any significant association between testing debt, productivity or quality~\cite{Klotins2018}.

Statistical results showed a significant association between start-up team size, team skill level and level of technical debt. Larger, less skilled teams are more likely to suffer from consequences of technical debt ($Cramer's~V = 0.386, p < 0.05$).

Comparing reports on the technical debt between start-ups at different life-cycle
stages we found a pattern of architecture debt growing over time and peaking at
the growth stage (C7). By looking into respondent reflections, we found that start-ups at growth and maturity stages face challenges stemming from earlier architecture decisions. As stated by one respondent:

\myquote{Initially, the product was started by an inexperienced developer and many design decisions were incorrect. Fixing them early would have been easy, however with the product and user base growing, 
changing core things became increasingly difficult. Correcting simple architecture mistakes that could 
have been trivial to fix early, can take weeks now.}

Our results match earlier findings suggesting that technical debt in start-ups
is caused by a need to develop a product fast and an inadequate team skill level~\cite{Unterkalmsteiner}. However, our respondent answers indicate that external pressures, such as competition do not cause the need for speed. Instead, the need for fast product development is primarily caused by internal considerations to validate the product idea as quickly as possible with minimal waste.

\subsubsection{Practices}

Start-ups report leveraging on best practices and established frameworks in
creating software architectures (P11), thus avoiding the effort of inventing
complex solutions themselves. Nearly all start-ups report using open source or
free to use tools and components. Only few report use of commercial
off-the-shelf components. Such a strategy reduces the effort of software engineering, potential lock-in effects with specific vendors, and need to reinvent already existing functionality. As stated by one respondent:

\myquote{We used a standard MVC architecture for the back-end and the web front-end, and the standard architecture for iOs/Android. The back-end and web front-end was implemented with Ruby on Rails and we followed its architecture conventions to guide the design.}


Few start-ups in the maturity stage, 4 out of 15, 27\%, report using in-house
developed, innovative technologies. For example, highly customized
deployment configurations, audio and video processing tools, and a custom graph
database, to support their product development. Note that such innovations in
technologies are done late in the start-up life-cycle to support further
evolution of an already established product.

Graphical user interface design is interrelated with user experience design. User interfaces could be the primary touch point between users and the product, thus largely contributing to users perception of the product. In our study, we explore on how are the product user interface designs created without inferring a connection to a broader practice of user experience design.

Respondent responses on user interface (UI) design practices suggest that
start-ups use mock-ups, utilize design frameworks and borrow ideas from other
products to develop their product UI (P12). Three start-ups report outsourcing
the UI design work. Companies reflect that user interface design is an ongoing
process of continuous improvement. Thus, user interface design is not a significant concern
at any particular stage. At the inception and stabilization stages, UI is tested
by internal reviews and by observing users interacting with the UI. At later
stages of the start-up life-cycle when a large number of customers are interacting with the product, start-ups use analytical tools and A/B tests to monitor how customers interact with the UI and to identify opportunities for improvement.

The main reported goals of the UI are usability, 75\%, 63 cases, and usefulness, 35\%, 29 cases. When reflecting on their user interface development practices, respondents suggest that investing more time in customer tests instead of inventing the UI internally, and having clear user interface guidelines, would have been helpful.

Our results on user interface design are aligned with studies investigating user experience practices in a start-up context. Start-ups use a set of practices to iterate prototypes, collect feedback and continuously improve their user interfaces.~\cite{hokkanen2015ux}. Quality criteria for good user interfaces and user experience are a clear message, visual attractiveness, intuitiveness, and credibility~\cite{hokkanen2016minimum}.

\subsubsection{Lessons learned}

Our analysis contradicts earlier results that start-ups extensively use cutting-edge technologies~\cite{Giardino2014}. Rather, start-ups opt for a stable technology platform with standard architecture to alleviate technology risks. We formulate the following implications for practitioners:

\begin{itemize}
  \item Start-ups mitigate technology risks by selecting stable technologies and
    following best practices associated with the technologies. Stable
    technologies and best practices improve team performance and provide a rigid platform for innovative features.

  Further reading: Petersen et al.~\cite{petersen2017choosing} explore how companies make decisions to select between open-source, in-house, or COTS components. Although the study is performed among established companies, it presents different strategies and factors influencing the decisions, thus providing an overview of the process.
  
  Baskerville et al.~\cite{Baskerville2003} present a study exploring reactive engineering practices in start-up companies and argue that software engineering in start-ups mainly focuses on integration and interoperability of components developed elsewhere.

  \item Our findings suggest that user interface design is a significant activity. However, it must be done by continuously monitoring and tweaking the product.

  Suggested reading: Hokkanen~\cite{hokkanen2016minimum} presents a framework for minimum viable user experience pinpointing essential qualities and practices of developing good user interfaces.

\end{itemize}

\subsection{Project management} \label{sec_results_pm}
The project management process area concerns planning and control of engineering activities in a start-up. Planning and control are known as important to optimize resource usage and attainment of specific goals~\cite{snyder2014guide}. We aim to explore how start-ups plan and control their activities.

\subsubsection{Goals}
The start-up life-cycle model, see Fig.~\ref{fig_states}, outlines the main milestones, releasing a minimum viable product, stabilizing the product for growth, attaining market share, and transitioning into an established organization. 
We inquired start-ups on how they control and measure their progress towards success. 

The responses show that start-ups primarily use external metrics, such as revenue, number of customers, and customer
satisfaction, to measure their success (G15, P13). However, start-ups at growth and maturity stages also consider using internal metrics, such as team performance, adherence to deadlines, and budget plans as performance measures
(G16, P14).

We observe differences between start-ups in different life-cycle stages. Start-ups at the inception stage, before they have launched their product, do not report the use of any metrics to measure their progress. However, they aim to use external metrics, e.g., number product users, as soon as the product is launched. 
At the stabilization stage, just after the product is launched, the primary success metrics are external and aimed to monitor general product adoption rates.
As start-ups progress through the life-cycle, metrics become more specific, attached to high-level business milestones, and consider both internal and external aspects jointly. Internal team performance metrics are monitored and used to gauge start-up performance, in addition to external, market adoption metrics.

\subsubsection{Challenges}
\label{section_pm_challenges}

Our results show that start-ups aim to use internal and external metrics to gauge their progression, see G15-16 in Fig.~\ref{fig_tpa-overview}. However, external metrics are not available in the inception stage, before the product is launched. 

Start-ups at the inception stage do not report using any specific metrics to control their progress. Potentially, due to unclear, changing plans, and immature project management practices. Thus, start-ups at the inception stage lack a benchmark to gauge their progression (C8).
As stated by one practitioner:

\myquote{We work towards the goal to improve our platform, priorities changed, and but there was no control over schedule. Sometimes tasks took a lot longer than planned.}


In their reflections, practitioners mention a need for tighter control over product engineering work, deadlines and budget. As one practitioner reflected on lessons learned:

\myquote{First and foremost, I would have defined short term and long term goals of the start-up. That would have helped us to align all resources and activities.}

Results from the statistical analysis show that with the advantage of hindsight,
practitioners estimate the performance of their teams more critically by a
significant margin. Such findings suggest that it is challenging to assess performance from within a company objectively, and objective early performance indicators could be useful (C8). Objective performance indicators could be helpful to establish a performing team (C2).

Setting goals and metrics to measure the progress towards goals is a common practice in project management~\cite{snyder2014guide,Shahin2007}. However, as shown by practitioner responses, start-ups at the inception stage do not measure their progression. Lack of measurement could lead to pitfalls such as the late realization of resource overruns, and scope creep. 

Looking at the literature, we found several models aimed to guide early start-ups~\cite{Ries2011, Bosch2013,Blank2013b}. These models propose first to define and validate the customer need, then validate a solution to this need, followed by validating product feasibility through small and large-scale prototypes. A somewhat similar approach focused on continuous experimentation is proposed by Fagerholm~\cite{Fagerholm2017}. However, to what extent such practices are used to guide work in early start-ups remains to be explored.

Respondents estimate that their goals are rather clear and change rarely~\sparkline{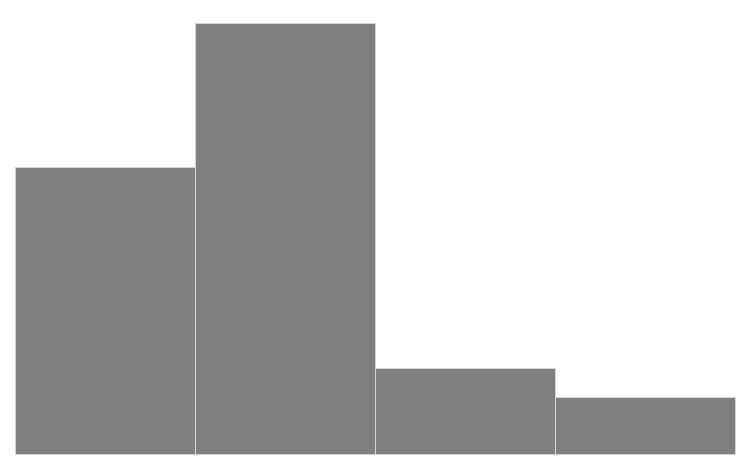}{To what extent it is true that goals in the start-up are rather unclear and change often?}. As the source for uncertainty respondents report specific requirements from customers, demands from partners, and commercial results (C9). Such factors are out of control for start-ups. Thus, any plans should have a room for uncertainty and adjustments.

When inquired to what extent respondents experience time and financial resources shortages,
most responses fall between ``A little'' and ``Somewhat'', indicating that availability financial resources and time is a concern, however not to an extreme extent, ~\sparkline{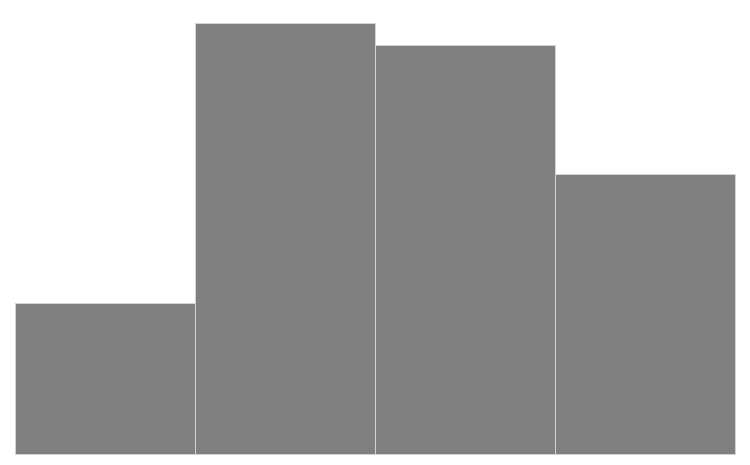}{To what extent it is true that there is a constant time pressure?} and \sparkline{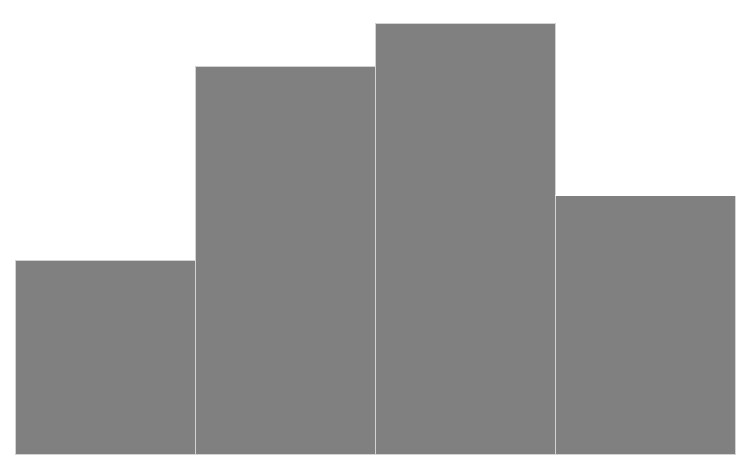}{To what extent it is true that there is a constant resources pressure?}

Statistical analysis shows a linear correlation between time and resources
shortages indicating that companies struggling with finances also struggle with
time. However, the free-text answers does not contain any mention of
specific difficulties stemming from time or financial resources shortages. The responses are consistent across the sample, and we could not identify any particular cohort where time and resource shortages would be more (or less, thereof) of a challenge.

Such results contradict earlier studies stating that extreme lack of resources
is one of the key characteristics of start-ups~\cite{Giardino2014,Sutton2000}. The
trade-off between project scope and budget, and project management are
reported as a concern in nearly every study exploring software project success
factors, for example, Chow et al.~\cite{Chow2008b}, Junk~\cite{Junk2000},
Reel~\cite{reel1999critical}, and Nasir~\cite{nasir2011critical}. Thus, time and
budget constraints, and associated trade-offs are not specific to start-ups exclusively. Moreover, a majority of software projects require more resources to
complete than initially estimated~\cite{molokken2003review}. Therefore, resource
shortages alone are not a differentiating factor between start-ups and established companies. That said, start-ups could differ
from established organizations with thinner margins for resource
overruns, less rigorous control over resource utilization, and could have more difficulty to access additional resources~\cite{bocken2015sustainable}.


A study of 1000 Finnish start-ups between 2010 – 2013 reveals that external funding has no association with start-up outcome. That is, start-ups with more resources at hand do not have more chances of survival and success than start-ups with more limited resources. Moreover, start-ups without external funding are likely to generate more revenue in the long term~\cite{Suominen2017}. Such findings highlight the importance of scoping, planning and control to optimize utilization of any amount of resources.
lack of clear goals


\subsubsection{Practices}


The responses suggest that start-ups at the inception stage use elementary, if any, practices to schedule their work and keep control over time and budget, see C8. Some start-ups are constrained with a fixed budget or a hard deadline, however control over utilization of time and budget is based on ``gut-feeling''. As one respondent from a start-up at the inception stage reflects on their planning practices:

\myquote{We have something on the budget but it's very informal and it's solely based on projected product revenue, and user retention and acquisition. In hindsight, we should have done better in terms of scheduling and scoping the work.}

Following from our earlier discussion on control over resource utilization, lack of control over time and financial resources is a potential pitfall for early start-ups.


Start-ups at the stabilization stage report improving project management practices, such as setting boundaries for certain expense positions, assigning a budget to meet specific goals, and attempting to estimate their cash-flow (P15). 
Start-ups at maturity stage report using processes, supported by tools, for resource planning and control over resource utilization (P16). As one respondent reflects on their planning practices in a mature start-up:
\myquote{We initially created a budget plan to understand if our business is viable at all and to apply for a loan. Nowadays we use budget estimates as a tool for planning.}

\subsubsection{Lessons learned}




Our analysis shows that clear goals, control over resources and schedule are essential to track progress over start-up life-cycle. However, there is a challenge to objectively assess own performance before a product is launched to market. Lack of planning and control could contribute to budget and schedule overruns leading to wasted opportunities.

Lack of resources does not have a substantial effect on engineering practices and start-ups' prospects of advancing through the life-cycle. However, lack of planning and control over resource utilization, especially on early stages of a start-up is a pitfall. Start-ups need to set clear goals and metrics to assess their performance from the very beginning objectively.

Suggested reading: Dybå et al.~\cite{dybaa2014agile} present an overview of agile project management practices and provide practical tips on how to organize project management in uncertain and changing environments.


\section{Discussion}

\subsection{Reflections on the research questions}


Our primary research question explores what engineering patterns, that is, common goals, practices, challenges, contextual factors, can be ascertained in start-up companies. 

To identify the patterns, we use the start-up life-cycle model. The model 
supported clustering of cases by product life-cycle stage and outcome. Thus, 
observing how engineering focus changes over start-up life-cycle and enabling a 
discussion of what practices contribute to desirable state transitions. 
Furthermore, with the model and our results, we aim to provide a blueprint for 
studying start-ups as dynamic and multi-faceted entities. Such blueprint could 
enable compatibility of results from different studies, and contribute towards 
a coherent view of software engineering in start-ups.


In most process areas we observe evolving practices, from rudimentary at the early stages, to more mature in the latter stages. Such evolution regarding practice maturity is also noted by earlier studies, e.g., Giardino et al.~\cite{Unterkalmsteiner}. However, we present empirical results of specific practices, goals, challenges at each life-cycle stage, and discuss the results in the context of related work from similar contexts. This way we provide actionable support for practicing software engineering in start-ups.

Our results show that domain knowledge, technical expertise, and teamwork are the key components in the early stages of a start-up. Shortages of any of these components can be compensated with specific practices. For example, scarcity of domain knowledge in the team can be compensated with more rigorous requirements engineering practices helping to identify, acquire, and distribute information in the organization. Similarly, lack of technical expertise in the core team can be compensated by outsourcing engineering work to another team.

In the later stages, technical debt, growing team and lack of processes (e.g. 
quality assurance, project management, managing a large, distributed team) 
hinder engineering work. However, these challenges can be anticipated and 
addressed with appropriate practices.

\subsection{Evolution of software engineering practices in start-ups}

By looking at the patterns, we can identify the key concerns at each life-cycle stage.

At the inception stage, the most important concern is to assemble a small team of few individuals with sufficient domain knowledge, technical expertise and adequate financial resources to build the first release of a product. As our results show, different practices can compensate for shortcomings in domain knowledge (e.g., by more actively involving stakeholders and more rigorously documenting requirements), lack of technical expertise (e.g., by using external engineering team), and limited budget (e.g., by adjusting the scope of the release). 
Releasing the first version of a product depends on how efficiently a team can use their understanding of the domain and engineering expertise to produce software. Excessively large teams and difficult communication drains resources, time and motivation, and could lead to the closure of the company even before the launch of a product.

At the stabilization stage, we observe two main concerns. The first is related to team building and establishing an efficient team with defined areas of responsibility, trust, and coordination. After a start-up releases the first product version to customers, the company need to balance between developing new features and providing quality service for the existing customers. Aforementioned requires the team to be more diverse, handle a broader range of responsibilities and more stringent internal structure. The second concern is related to establishing a feedback loop with customers. Our results show an association between using customer feedback and technical and commercial success. Identifying early customers and involving them in the engineering work is one of the essential tasks in a start-up.

At the growth stage, we observe that concerns related to business, monetization, and marketing become relevant and have an influence on product engineering. For example, companies' own goals such as the need for monetization and expansion, must be taken into account in addition to customer needs. Serving a growing number of potentially diverse customers require the product to be flexible, scalable, and reliable. Providing such qualities require expert engineers. Growing complexity of the product creates a need for more robust testing and deployment practices, alleviating a need for manual work to prepare every release. Furthermore, increasing complexity of the organization and the product expose technical debt which must be addressed to support further growth.

At the maturity, stage start-ups are concerned with managing a growing, and potentially, distributed team. At this point, more rigorous project management practices are needed to organize and control engineering work, and more thorough engineering processes could be introduced marking a transition into an established organization.

\section{Conclusion}

In this paper, we investigated how 84 software start-ups utilize software 
engineering to build innovative software-intensive products. To frame the 
results we proposed start-up life-cycle model and looked into team, 
requirements engineering, value focus, quality and testing, architecture and 
design, and software project management process areas. This framing highlights 
stage and process area specific goals, challenges, and practices. We have 
discussed our findings in the context of related work and formulated practical 
lessons learned aimed at practitioners. 

We conclude that all explored process areas are relevant for start-ups and, in 
essence, not different from established companies. That said, potentially the 
key difference and difficulty to practice software engineering in start-ups is 
to manage the evolution of practices in all the process areas at once with only 
a little room for error. Even though start-ups are experimental by nature and making errors in the process is inevitable, there is a distinction between taking calculated risks, and neglecting the best engineering practices.

To further understand and support software engineering in start-ups we aim to conduct a series of structured workshops to conduct an assessment of software engineering practices in start-ups using results obtained in this study. With the results of the workshops, we aim to refine further the patterns and a methodology for using the patterns in improving engineering practices in start-ups.

\section{Acknowledgments}
The authors of this paper would like to thank all practitioners who found time 
and motivation to share their experiences. Reaching this diverse population of 
start-ups would not be possible without help and support from Software Start-up 
Research Network\footnote{The Software Start-up Research 
Network, 

https://softwarestartups.org/} community,
and specifically Nana Assyne, Anh Nguyen Duc, Ronald Jabangwe, Jorge Melegati, Bajwa Sohaib Shahid, Xiaofeng Wang, Rafael Matone Chanin, and Pekka Abrahamsson.

Work of Rafael Prikladnicki is partially funded by CNpq and FAPERGS (process 17/2551-0001205-4).

\ifCLASSOPTIONcaptionsoff
  \newpage
\fi


\bibliography{./library.bib}{}
\bibliographystyle{IEEEtran}

%


\begin{IEEEbiography}
    [{\includegraphics[width=66pt,clip,keepaspectratio]{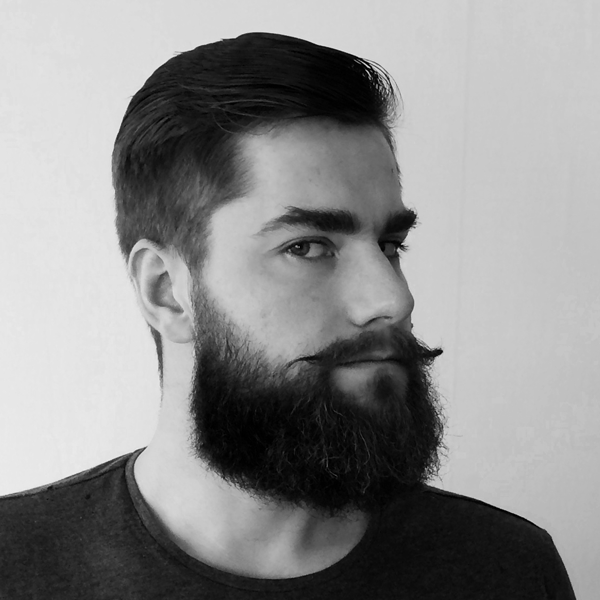}}]
    {Eriks Klotins} is a Ph.D. student of Software Engineering at Blekinge Institute of Technology (BTH). He received his Master's degree from University of Latvia in 2011. He has over nine years of experience in software engineering in both large companies and start-ups. His research is focused on innovative software-intensive product engineering in start-ups. Specifically, what engineering practices and methodologies are best suited for start-ups to support their evolution, and commercial and technical success.
\end{IEEEbiography}

\begin{IEEEbiography}
    [{\includegraphics[width=66pt,clip,keepaspectratio]{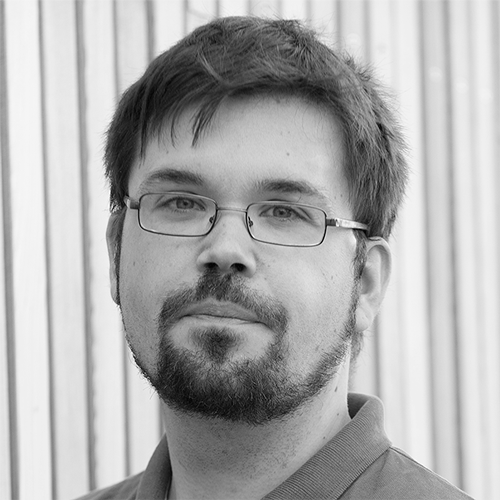}}]
    {Dr. Michael Unterkalmsteiner} is a senior lecturer in the Blekinge Institute
    of Technology’s Software Engineering Research Laboratory. His research
    interests include data-driven software engineering, software measurement
    and testing, process improvement, and requirements engineering. He
    primarily conducts empirical research in close collaboration with
    industry partners. Michael received a PhD in software engineering from
    the Blekinge Institute of Technology. 
\end{IEEEbiography}

\begin{IEEEbiography}
    [{\includegraphics[width=66pt,clip,keepaspectratio]{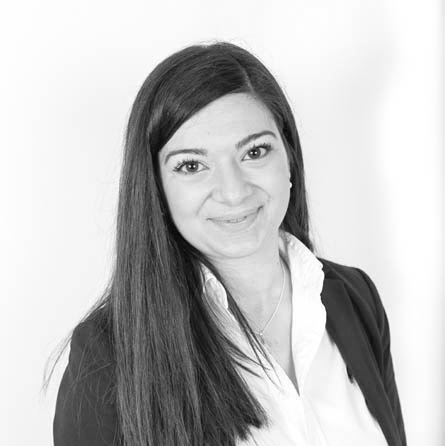}}]
{Dr. Panagiota Chatzipetrou} is an Assistant Professor at the department of Informatics at Örebro University School of Business in Örebro, Sweden, where she belongs to the Centre for empirical research on information systems (CERIS). She is also part of the Software Research Engineering Lab (SERL) at Blekinge Institute of Technology in Karlskrona, Sweden. She received her BSc degree in “Informatics”, MSc in “Informatics and Business Administration” and Ph.D. in “Informatics” from the Department of Informatics, Aristotle University of Thessaloniki (AUTh), Greece. Her doctoral dissertation has the title: “Statistical methods in information systems project planning”. She holds a master in pedagogy and didactics and in parallel she has been educated in special education, learning difficulties and dyslexia. As a researcher, she mainly focuses on empirical studies under the different perspectives of software development. Her research interests include - but are not limited to- applications of statistical methods to quality problems in software engineering and especially to requirements engineering and the exploitation of human factor and the different views that ultimately determine the quality of a software product and the product development. Also, she has been working with decision support systems for the development of software-intensive systems, large-scale agile (and global) software development, and behavioral software engineering.
\end{IEEEbiography}

\begin{IEEEbiography}
    [{\includegraphics[width=66pt,clip,keepaspectratio]{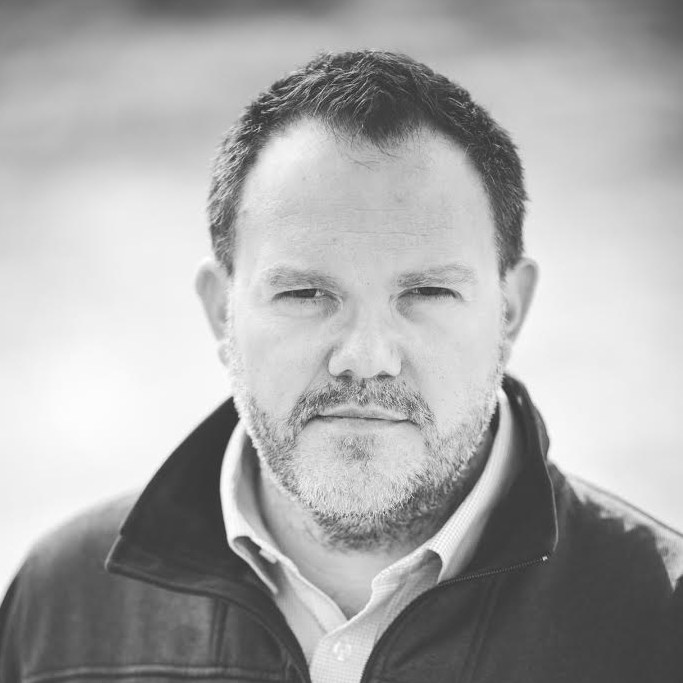}}]
    {Prof. Dr. Tony Gorschek} is a professor of Software Engineering at Blekinge Institute of Technology (BTH) and part time at Chalmers. He has over ten years industrial experience as a CTO, senior executive consultant and engineer, but also as chief architect and product manager. In addition he has built up five startups in fields ranging from logistics to Internet based services.
\end{IEEEbiography}

\begin{IEEEbiography}
    [{\includegraphics[width=66pt,clip,keepaspectratio]{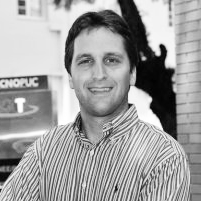}}]
    {Rafael Prikladnicki} is an associate professor at the School of Technology and the director of the Science and
Technology Park (Tecnopuc) at the Pontifical Catholic University
of Rio Grande do Sul, Brazil, where he also leads the MuNDDoS research
group. He is the chair of IEEE Software's advisory board. 
\end{IEEEbiography}

\begin{IEEEbiography}
  [{\includegraphics[width=66pt,clip,keepaspectratio]{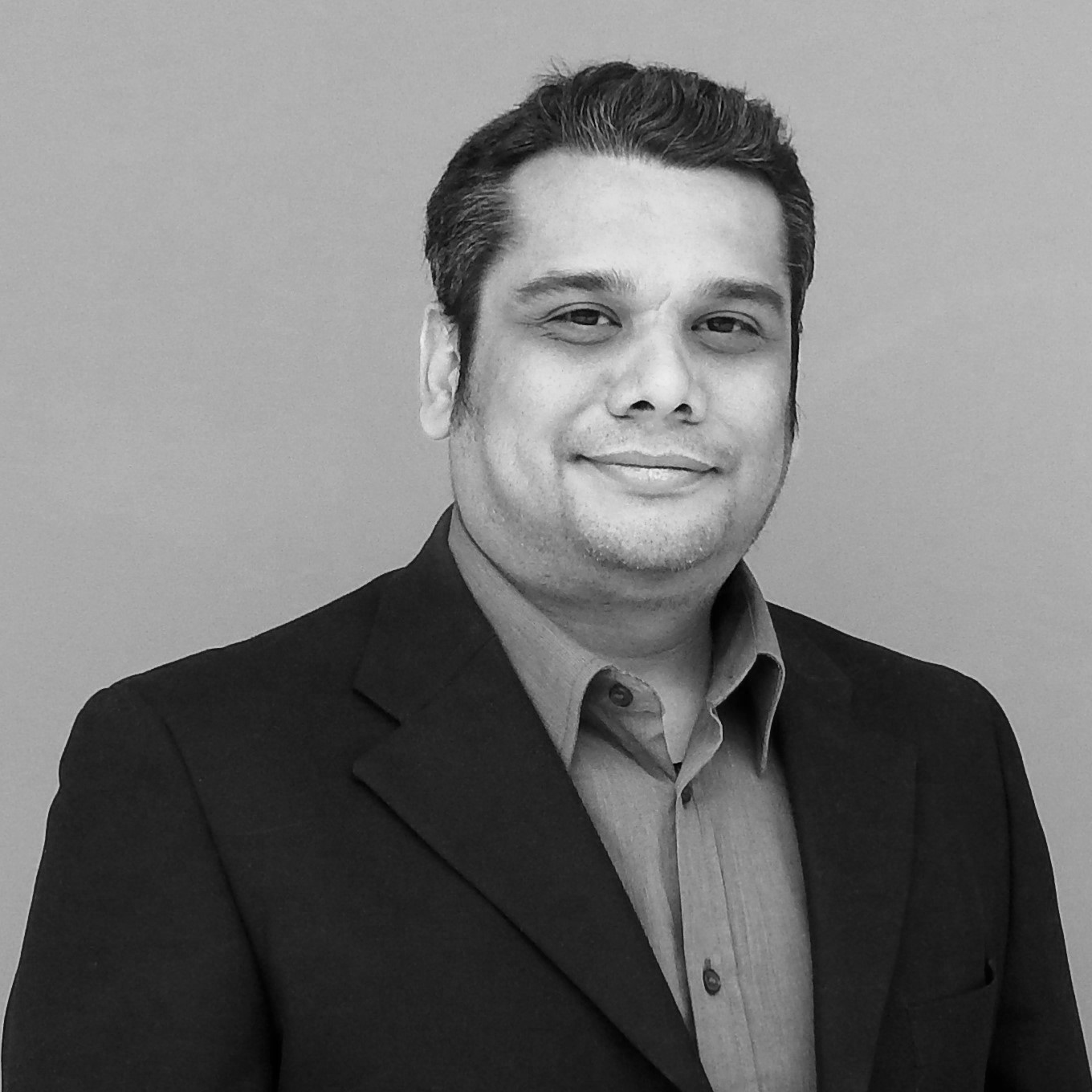}}]
  {Nirnaya Tripathi} is a doctoral researcher at M3S research unit, University of Oulu.  He received his Master’s degree in Information Processing Science at the University of Oulu in 2012.  Before that, he received his bachelor’s degree in Information Technology at Manipal Institute of Technology in 2009. His research interest includes entrepreneurship, startup ecosystem, software engineering in startup, large-scale lean and agile software development and global software development.  He is currently doing research on startup ecosystem, minimum viable product, and requirements engineering in software startups.
\end{IEEEbiography}

\begin{IEEEbiography}
   [{\includegraphics[width=66pt,clip,keepaspectratio]{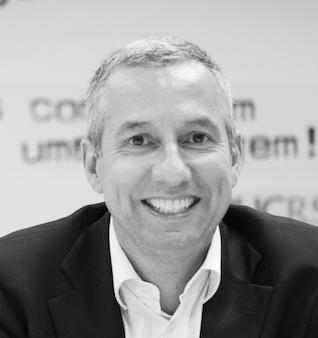}}]
   {Leandro Bento Pompermaier} is an assistant professor at the School of Technology at the Pontifical Catholic University of Rio Grande do Sul (PUCRS) - Brazil. In addition, he is the leader of the startup area of the PUCRS Science and Technology Park (TECNOPUC). He holds a master degree in computer science from the Federal University of Rio Grande do Sul and is currently a PhD student at PUCRS. He is also an entrepreneur and angel-investor in some IT companies and software startups.
\end{IEEEbiography}







\end{document}